\documentclass[11pt]{article}
\usepackage{amssymb}

\newcommand\unit{\hbox{\rm 1\kern-2.8truept l}}
\newcommand\re{{\Re}\kern-1pt e}
\newcommand\im{{\Im}\kern-1pt m}
\newcommand\tr{\hbox{\rm tr}}
\newcommand{\B}{\mathcal{B}(\h) }

\newcommand{\qed}{\hfill (q.e.d.)} 
\newcommand{\wTs}{\widetilde{\mathcal{T}}^{(s)}}
\newcommand{\wT}{\widetilde{\mathcal{T}}^{(0)}}
\newcommand{\wLs}{\widetilde{\mathcal{L}}^{(s)}}
\newcommand{\wL}{\widetilde{\mathcal{L}}^{(0)}}
\newcommand{\tL}{\widetilde{\mathcal{L}}}
\newcommand{\tT}{\widetilde{\mathcal{T}}}
\newcommand{\anal}{\!\textit{\Large{a}}\, }

\newcommand{\ee}[2]{\mid\! #1\rangle\langle #2\!\mid}
\newcommand\Lform{{\mathcal{L}}\kern-7.56pt\raise1.0pt\hbox{$-$}}

\newcommand{\h}{\mathsf{h}}
\newcommand{\dimo}{\noindent{\bf Proof.\ }}
\newcommand{\T}{\mathcal{T}}
\newcommand{\Ll}{\mathcal{L}}
\newcommand{\ketbra}[2]{\mid\! #1\rangle\langle #2\!\mid}
\newtheorem{definition}{Definition}
\newtheorem{theorem}[definition]{Theorem}
\newtheorem{proposition}[definition]{Proposition} 
\newtheorem{lemma}[definition]{Lemma}

\newtheorem{remark}[definition]{Remark}
\newtheorem{example}[definition]{Example}

\begin{document}

%\markboth{Authors' Names}{Paper's Title}

\title{\bf GENERATORS OF DETAILED BALANCE QUANTUM MARKOV SEMIGROUPS }

\date{}

\maketitle

\centerline{\author{FRANCO FAGNOLA and VERONICA UMANIT\`A}}\vskip 0.2cm

\centerline{\small{Dipartimento di Matematica, Politecnico di Milano,}} 
\centerline{\small{Piazza Leonardo da Vinci 32, I-20133 Milano (Italy)}}
\vskip 0.2cm
\centerline{\small{franco.fagnola@polimi.it,}}
\centerline{\small{veronica.umanita@fastwebnet.it}}

\begin{abstract}
For a quantum Markov semigroup $\T$ on the algebra $\B$ with a faithful invariant 
state $\rho$, we can define an adjoint $\widetilde{\T}$ with respect to the scalar product 
determined by $\rho$. In this paper, we solve the open problems of characterising  
adjoints $\widetilde{\T}$ that are also a quantum Markov semigroup and satisfy the 
detailed balance condition in terms of the operators $H,L_k$ in the Gorini 
Kossakowski Sudarshan Lindblad representation $\Ll(x)=i[H,x] - 
\frac{1}{2}\sum_k(L^*_kL_k x-2L^*_kxL_k + xL^*_kL_k)$ of the generator of $\T$. 
We study the adjoint semigroup with respect to both scalar products 
$\langle a,b\rangle = \tr(\rho a^* b)$ and $\langle a,b\rangle 
= \tr(\rho^{1/2} a^* \rho^{1/2}b)$.
\end{abstract}

\noindent
\small{Kerwords: Quantum detailed balance, quantum Markov semigroup, Lindablad representation}
\vskip 0.2cm


\section{Introduction} 

The principle of detailed balance is at the basis of equilibrium physics. The notion 
of detailed balance for open quantum systems (Alicki\cite{alicki}, Frigerio and 
Gorini\cite{FrGo}, Kossakowski, Frigerio, Gorini and Verri\cite{kossa}, Alicki and 
Lendi \cite{AlickiLendi}) when the evolution is described by a uniformly continuous 
Quantum Markov Semigroup (QMS) $\T$ with a faithful normal invariant state $\rho$, 
is formulated as a property of the generator $\Ll$. 
Indeed, for a system with associated separable Hilbert space $\h$, this 
can be written in the Gorini Kossakowski Sudarshan Lindblad (GKSL) form 
\begin{equation}\label{eq-GKSL}
\Ll(x)=i[H,x]-\frac{1}{2}\sum_k(L^*_kL_k x-2L^*_kxL_k+ xL^*_kL_k)
\end{equation}
where $H$ and $L_k$ are operators on $\h$ that can always be chosen satisfying 
$\tr(\rho L_k)=0$ and the natural summability and minimality conditions 
(see Theorem \ref{th-special-GKSL}).
The state $\rho$ defines a scalar product 
$\langle x, y\rangle = \tr(\rho x^*y)$ on the algebra 
$\mathcal{B}(\h)$ of operators on $\h$, and $\mathcal{T}$ admits a dual semigroup with 
respect to this scalar product if there exists another  uniformly continuous QMS $\tT$ 
(generated by $\tL$) such that 
$\tr(\rho \tT_t(x)y)=\tr(\rho x\mathcal{T}_t(y))$. The QMS $\T$ satisfies 
the quantum detailed balance condition if the effective Hamiltonian $H$ commutes 
with $\rho$ and $\widetilde{\Ll}=\Ll-2i[H,\cdot]$ i.e. the dissipative part 
$\Ll_0=\Ll-i[H,\cdot]$ of $\Ll$ is self-adjoint.

This generalizes the notion of detailed balance (reversibility) for a classical 
Markov semigroup which is called reversible when it is self-adjoint in the $L^2$ 
space of an invariant measure. It is worth noticing, however, that, in the commutative 
case, the adjoint (dual) of a Markov semigroup is always a Markov semigroup. 

The dual of a QMS $\T=(\T_t)_{t\ge 0}$ with respect to the state 
$\rho$ may not be a QMS because the adjoint $\widetilde{\T}_t$ of the map $\T_t$ 
may not be positive or, even more, may not be a $^*$-map i.e. 
$\widetilde{\T}_t(a)^*\not=\widetilde{\T}_t(a^*)$. It is known (see e.g. 
Ref.\cite{kossa} Prop. 2.1) that $\widetilde{\T}_t$ is a completely positive 
map if it commutes with the modular group $(\sigma_t)_{t\in\mathbb{R}}$ 
($\sigma_t(a)=\rho^{it}a\rho^{-it}$) associated with $\rho$. More recently, 
Majewski and Streater\cite{maje} (Thm.6 p.7985) showed that the $\widetilde{\T}_t$ 
are (completely) positive whenever they are $^*$-maps.

The structure of the generator (\ref{eq-GKSL}) of a detailed balance QMS was 
studied in Ref.\cite{alicki} and Ref.\cite{kossa} under the additional assumption 
that it is a normal operator, i.e. $\Ll$ and $\widetilde{\Ll}$ commute.

In this paper, we solve the open problems of characterising in terms of 
the operators $H, L_k$ in (\ref{eq-GKSL}) dual semigroups $\widetilde{\T}$ 
that are QMSs and when they satisfy the detailed balance condition without 
additional assumptions on the generator $\Ll$. The main results of this 
paper, Theorems \ref{HLtilde} and \ref{th-HL-0-DB}, describe the structure 
of generators of QMS whose dual is still a QMS and, among them, the 
structure of those satisfying the detailed balance condition.

The dual semigroup is a QMS if and only if the maps $\T_t$ 
commute with the modular automorphism (Theorem \ref{th-dual-QMS-impl-commautmod}). 
When this happens we can find particular GKSL representations of $\Ll$ as in 
(\ref{eq-GKSL}), that we call privileged, with $H$ commuting with $\rho$ and 
the $L_k, L_k^*$'s eigenvalues of the modular automorphism (Definition 
\ref{def-privileged-GKSL}) i.e. $\rho L_k\rho^{-1}=\lambda_k L_k$. 
Moreover, the generator $\widetilde{\Ll}$ of the dual semigroup admits a 
privileged GKSL representation with $\widetilde{H}= - H - c$ ($c$ 
is a real constant) and $\widetilde{L}_k=\lambda_k^{-1/2}L^*_k$ (Theorem \ref{HLtilde}).

Finally the quantum detailed balance condition $\Ll-\widetilde{\Ll}=2i[K,\cdot]$ 
(for some self-adjoint operator $K$) holds if and only if $H=K+c$ and there 
exists a unitary matrix $(u_{k\ell})_{k\ell}$ such that $\lambda_k^{-1/2}L^*_k
= \sum_\ell u_{k\ell}L_\ell$ (Theorem \ref{th-HL-0-DB}).

There are other choices of the scalar product on ${\mathcal{B}}(\h)$ induced 
by $\rho$; we can define $\langle a,b\rangle_s=\tr(\rho^{1-s}a^*\rho^s b)$ for 
any $s\in[0,1]$. The most studied case is the previous one with $s=0$. 
The case $s=1/2$, sometimes called {\em symmetric}, however, is also interesting
(see Goldstein and Lindsay\cite{GoLi}). 
Indeed, as wrote Accardi and Mohari\cite{AcMo} (p.409), ``it is worth characterizing 
the class of Markov semigroup such that $\T_t=\widetilde{\T}_t$'' in full 
generality also for the dual semigroup with respect to the ``symmetric'' scalar 
product $\langle a,b\rangle=\tr(\rho^{1/2}a^*\rho^{1/2}b)$ (Petz's duality). 
Note that the ``symmetric'' dual semigroup is always a QMS.

In Section \ref{sect-symmBD} we solve this problem and the more general 
problem of characterising QMSs satisfying the ``symmetric'' detailed balance 
condition $\Ll-\Ll^\prime=2i[K,\cdot]$ ($\Ll^\prime$ is the generator of the 
symmetric dual QMS). We show that the ``symmetric'' detailed balance is 
weaker than usual detailed balance (Proposition \ref{1/2impl0}) and establish 
the relationships among the $L_k$'s, the operator $G=-2^{-1}\sum_k L^*_kL_k-iH$
and $\rho$ of symmetric detailed balance $\Ll$ (Theorem \ref{th-gen-symm-DB}). 
Examples \ref{example-H-no-comm-rho-dual} and \ref{example-H-no-comm-rho-symm-DB}
show that, in the ``symmetric'' case, the effective Hamiltonian $H$ may not 
commute with $\rho$.

The paper is organised as follows. In Section \ref{sect-class-DB} we outline 
the detailed balance condition for classical Markov semigroups. Then we 
explore several possible definitions of the dual semigroup in Section 
\ref{sect-quantum-dual} and study the generators of QMS whose dual is 
still a QMS in Section \ref{sect-generator-quantum-dual}. In Section 
\ref{sect-Q-DB} we characterise generators of quantum detailed balance 
QMS. The special case of QMS on $2\times 2$ matrices is analysed in 
Section \ref{sect-QMS-onM2C}; in this case it turns out that, if the 
dual semigroup is a QMS, then it satisfies the quantum detailed balance 
condition. Further examples of detailed balance QMSs, also with unbounded 
generators can be found in the literature and in Ref.\cite{FagQue}. 
Finally, in Section \ref{sect-symmBD}, we study the symmetric 
detailed balance condition.

% relazioni con lavori di Barndorff-Nielsen et al. riguardo a struttura di certi 
% Lindbladiani in misurazioni continuate? 

% progetto futuro di paragonare T e il suo duale nel non-equilibrio

\section{Classical detailed balance}\label{sect-class-DB}
Let $(E,\mathcal{E},\mu)$ be a measure space with $\mu$ $\sigma$-finite 
and let $T=(T_t)_{t\ge 0}$ be a weakly* continuous Markov semigroup of bounded 
positive linear maps on $L^\infty(E,\mathcal{E},\mu)$. $T$ is the dual 
semigroup of a strongly continuous contraction semigroup on the predual space 
$L^1(E,\mathcal{E},\mu)$ denoted $T_*$. Suppose that $T$ admits
a $T$-invariant probability density $\pi$ (a norm one, non-negative, 
function in $L^1(E,\mathcal{E},\mu)$ such that $T_{*t}\pi=0$ for $t\ge 0$) 
vanishing only on an element of $\mathcal{E}$ of measure $0$. Then, it is 
well-known that the sesquilinear form
\[
(f,g) =\int_E\overline{f}g\pi\ d\mu
\] 
defines a scalar product on $L^\infty(E,\mathcal{E},\mu)$, that 
we denote by $\langle\cdot,\cdot\rangle_{\pi}$, and putting 
\begin{equation}\label{formula}
\widetilde{T}_t(g)=\pi^{-1}T_{*t}(\pi g)
\end{equation} 
for each $t\geq 0$ one defines the adjoint of the operator $T_t$ with respect 
to this scalar product. Indeed, $\pi g$ belongs to $L^1(E,\mathcal{E},\mu)$ and the 
$T_*$-invariance of $\pi$ yields
\[
\left|\widetilde{T}_t(g)\right|\leq\pi^{-1}T_{*t}(\pi)
\Vert g\Vert_\infty=\Vert g\Vert_\infty,
\]
so that $\widetilde{T}_t$ is a well defined bounded operator on
$L^\infty(E,\mathcal{E},\mu)$. Moreover, we have
\begin{eqnarray*}
\langle\widetilde{T}_t(g),f\rangle_\pi
&=&\int_E\overline{\widetilde{T}_t(g)}f\pi
d\mu=\int_E\pi^{-1}\overline{{T}_{*t}(\pi g)}f\pi d\mu\\
&=&\int_E {T}_{*t}(\pi \overline{g})f
d\mu=\int_E(\pi\overline{g})T_t(f)\, d\mu(x)
=\langle g,T_t(f)\rangle_\pi
\end{eqnarray*} 
for every $f,g\in L^\infty(E,\mu)$.
Clearly the maps $\widetilde{T}_t$ are also positive, thus 
$\widetilde{T}=(\widetilde{T}_t)_{t\geq 0}$ is a weakly* continuous 
semigroup of bounded positive maps on $L^\infty(E,\mathcal{E},\mu)$.

Finally, the semigroup $\widetilde{T}$ is Markov since 
$\widetilde{T}_t(\unit)=\pi^{-1}T_{*t}(\pi)=\unit$.

\begin{definition} We say that $T$ satisfies \emph{classical detailed 
balance} if every operator $T_t$ is selfadjoint with respect to 
$\langle\cdot,\cdot\rangle_{\pi}$, i.e. $\widetilde{T}_t=T_t$.
\end{definition}

Therefore, $T$ satisfies classical detailed balance if and only if 
\begin{equation}\label{eq-class-Markov-semigroup-det-bal}
{T}_t(f)=\pi^{-1} {T}_{*t}(\pi f).
\end{equation}

\begin{remark}\rm Detailed balance is equivalent to reversibility of classical 
Markov chains. Indeed, when $E=\{\,1,\dots,d\,\}$ is a finite (for simplicity) 
set, endowed with the discrete $\sigma$-algebra $\mathcal{E}$ and the counting 
measure $\mu$, with a Markov semigroup $(T_t)_{t\ge 0}$ we can associate the 
transition rate matrix $(q_{jk})_{1\le j,k\le d}$ defined by
\[
q_{jk} = \lim_{t\to 0}t^{-1}\left(T_t 1_{\{k\}}-1_{\{k\}}\right)(j)
\]
($1_{\{k\}}$ denotes the indicator function of the set ${\{k\}}$). Denoting 
$(\widetilde{q}_{jk})_{1\le j,k\le d}$ the transition rate matrix associated with 
the Markov semigroup $(\widetilde{T}_t)_{t\ge 0}$, it follows immediately 
from the definitions that (\ref{eq-class-Markov-semigroup-det-bal}) is equivalent 
to the classical condition $\pi_j{q}_{jk}=\pi_k q_{kj}$ for all $j,k\in E$ 
called {\em reversibility}.

The same condition also arises in discrete time Markov chains.
\end{remark}

\section{The quantum dual semigroup}\label{sect-quantum-dual}

The definition of detail balance involves the dual semigroup with respect 
to the scalar product determined by the invariant state. When studying the 
non-commutative analogue two fundamental differences with the classical commutative 
case arise: 1) there are several possible dualities, 2) the dual semigroup might 
not be positive. In this section we analyze these problems.

Let $\h$ be a complex separable Hilbert space and let $\mathcal{T}$ 
be a uniformly continuous QMS on $\mathcal{B}(\h)$ generated by a bounded linear 
operator $\mathcal{L}$. A faithful invariant state $\rho$ for $\mathcal{T}$ can 
be written in the form
\begin{equation}\label{eq-rho}
\rho=\sum_{k\geq 1}\rho_k |e_k\rangle\langle e_k|,
\end{equation}
where $\rho_k>0$ for every $k$, $\sum_{k\geq 1}\rho_k=1$ and 
$(e_k)_{k\ge 1}$ is an orthonormal basis of $\h$. Therefore $\rho$ is invertible but, when $\dim\h=\aleph_0$, its inverse $\rho^{-1}=\sum_{k\geq 1}\rho_k^{-1}\ee{e_k}{e_k}$ 
is a positive operator with dense domain $\rho(\h)$.

\begin{definition}
Let $s\in[0,1]$ fixed. We say that $\mathcal{T}$ admits the \emph{$s$-dual semigroup} with respect to $\rho$ if there exists a uniformly continuous semigroup $\widetilde{\mathcal{T}}=\{\widetilde{\mathcal{T}}_t\}_t$ on $\B$ such that
\begin{equation}\label{s-DB}
\tr(\rho^{1-s}\widetilde{\mathcal{T}}_t(a)\rho^s b)
=\tr(\rho^{1-s}a\rho^s \mathcal{T}_t(b))
\end{equation} 
for all $a,b\in\B$, $t\geq 0$.\\
When $s=0$ we shall abbreviate the name of $\widetilde{\mathcal{T}}$ speaking of dual semigroup.\\
We denote by $\mathcal{T}_{*t}$ and $\widetilde{\mathcal{T}}_{*t}$ the 
predual maps 
of $\mathcal{T}_{t}$ and $\widetilde{\T}_t$ respectively.\\     
\end{definition}
We remark that for every $s\in[0,1]$ the sesquilinear form
\[
\langle a,b\rangle_s:=\tr(\rho^{1-s}a^*\rho^sb)
\] 
defines a scalar product on $\mathcal{B}(\h)$: indeed 
\[
\langle a,a\rangle_s = 
\tr((\rho^{s/2}a\rho^{(1-s)/2})^*(\rho^{s/2}a\rho^{(1-s)/2}))\ge 0
\]
and $\langle a,a\rangle_s=0$ implies $\rho^{s/2}a\rho^{(1-s)/2}=0$, 
i.e. $a=0$ because $\rho$ is invertible.\\
If $\widetilde{\mathcal{T}_t}$ is a $*$-map, then it is exactly the adjoint operator of $\mathcal{T}_t$ with respect to the scalar product $\langle\cdot,\cdot\rangle_s$.

In our framework, we will always suppose that $\mathcal{T}$ admits the $s$-dual semigroup.
\begin{proposition} \label{propTtilde} 
For each $t\geq 0$ and $a\in\mathcal{B}(\h)$ we have  
\begin{equation}\label{eq-Ttilde-T}
\rho^{1-s}\widetilde{\mathcal{T}}_t(a)\rho^{s}
=\mathcal{T}_{*t}(\rho^{1-s}a\rho^{s}).
\end{equation}
Moreover, the following properties hold:
\begin{enumerate}
\item $\widetilde{\T}_t({\hbox{\em 1\kern-2.8truept l}})={\hbox{\em 1\kern-2.8truept l}}$;
\item $\widetilde{\T}_{*t}(\rho)=\rho$;
\item if $\widetilde{\T}_t$ is positive, then it is also normal.
\end{enumerate}
\end{proposition}

\dimo 
The identity (\ref{eq-Ttilde-T}) is easily checked starting from (\ref{s-DB}) and 
using that \par\noindent $\tr(\rho^{1-s}a\rho^{s}\mathcal{T}_{t}(b))$ $=\tr(\mathcal{T}_{*t}(\rho^{1-s}a\rho^{s})b)$. \\
Putting $a=\unit$, we find then $\rho^{1-s}\widetilde{\mathcal{T}}_t(\unit)\rho^{s}
=\mathcal{T}_{*t}(\rho) = \rho$ by the invariance of $\rho$; this implies $(\widetilde{\mathcal{T}}_t(\unit)-\unit)\rho^s=0$, i.e. $\widetilde{\mathcal{T}}_t(\unit)=\unit$ for the density of $\rho(\h)$ in $\h$.\\
Taking $b=\unit$ in (\ref{s-DB}) yields
$\tr(\widetilde{\mathcal{T}}_t(a)\rho)=\tr(a\rho)$ for all $a\in\B$. This means in particular that the map $a\mapsto\tr(\widetilde{\mathcal{T}}_t(a)\rho)$ is weakly*-continuous, so $\rho$ belongs to the domain of $\widetilde{\mathcal{T}}_{*t}$ and $\widetilde{\mathcal{T}}_{*t}(\rho)=\rho$.\\
To prove property $3$ it is enough to show that, for every increasing net $(x_\alpha)_\alpha$ of positive elements in $\B$ with $\sup_\alpha x_\alpha=x\in\B$, we have 
$$
\lim_\alpha\langle u,\widetilde{\T}_t(x_\alpha)u\rangle=\langle u,\widetilde{\T}_t(x)u\rangle
$$
for each $u$ in a dense subspace of $\h$.\\
So, let $u\in\rho(\h)$; then $u=\rho^{1-s}v=\rho^sw$ for some $v,w\in\h$. Therefore, equation \ref{eq-Ttilde-T} implies
\begin{eqnarray*}
\lim_\alpha\langle u,\widetilde{\T}_t(x_\alpha)u\rangle&=&\lim_\alpha\langle v,\rho^{1-s}\widetilde{\T}_t(x_\alpha)\rho^sw\rangle
=\lim_\alpha\langle v,\T_{*t}(\rho^{1-s} x_\alpha\rho^s)w\rangle\\ 
&=&\langle v,\T_{*t}(\rho^{1-s} x\rho^s)w\rangle=\langle v,\rho^{1-s} \widetilde{\T}_t(x)\rho^sw\rangle=\langle u,\widetilde{\T}_t(x)u\rangle,
\end{eqnarray*}
since $\T_{*t}$ is normal.
\qed\smallskip\\

It is clear from (\ref{eq-Ttilde-T}) that 
\begin{equation}\label{deftildefindim}
\widetilde{\mathcal{T}}_t(a)
=\rho^{-(1-s)}\mathcal{T}_{*t}(\rho^{1-s}a\rho^{s})\rho^{-s}
\end{equation}
on the dense subset $\rho^s(\h)=\rho(\h)$ of $\h$, so that the $1/2$-dual semigroup is completely 
positive and then it is a QMS thanks to Proposition \ref{propTtilde}.3. However, for $s\not=1/2$, contrary to what happens in 
the commutative case, the maps $\widetilde{\mathcal{T}}_t$ might not be positive. 
In this case $\widetilde{\T}$ is not a QMS (see Example \ref{tildenoQMS}).
\begin{remark}\rm If $\h$ is finite-dimensional, then any uniformly continuous QMS $\T$ on $\B$ admits the $s$-dual semigroup, since equation \ref{deftildefindim} defines a uniformly continuous semigroup of bounded operators on $\B$ satisfing 
$\tr(\rho^{1-s}\widetilde{\mathcal{T}}_t(a)\rho^s b)
=\tr(\rho^{1-s}a\rho^s \mathcal{T}_t(b))$. 
\end{remark}
The relationships between the generators $\Ll$, $\widetilde{\Ll}$, 
${\Ll}_*$ and ${\widetilde{\Ll}}_*$, of $\T$, $\widetilde{\T}$, 
${\T}_*$ and ${\widetilde{\T}}_{*}$ respectively are easily deduced. 

\begin{proposition}\label{prop-s-db-generators}
The semigroups $\T$ and $\widetilde{\T}$ satisfy (\ref{s-DB}) if 
and only if, for all $a,b\in\mathcal{B}(\h)$, we have 
\begin{equation}\label{eq-s-db-generators}
{\hbox{\em tr}}(\rho^{1-s}\widetilde{\mathcal{L}}(a)\rho^s b)
={\hbox{\em tr}}(\rho^{1-s}a\rho^s \mathcal{L}(b)).
\end{equation} 
In this case, the following identity holds
\begin{equation}\label{eq-L-Ltilde}
\rho^{1-s}\widetilde{\mathcal{L}}(a)\rho^{s}=\mathcal{L}_*(\rho^{1-s}a\rho^{s}).
%,\qquad\rho^{s}\mathcal{L}(a)\rho^{1-s}=\widetilde{\mathcal{L}}_*(\rho^{s}a\rho^{1-s}). 
\end{equation}
Moreover, if $\widetilde{\mathcal{T}}$ is a QMS, then 
\begin{equation}\label{eq-L-Ltilde2}
\rho^{s}\mathcal{L}(a)\rho^{1-s}=\widetilde{\mathcal{L}}_*(\rho^{s}a\rho^{1-s})
\end{equation}
\end{proposition} 

\dimo The identity (\ref{eq-s-db-generators}) clearly follows 
differentiating (\ref{s-DB}) at $t=0$. Conversely, the identity 
(\ref{eq-s-db-generators}), implies that, for all $n\ge 0$ we have
\[
\tr(\rho^{1-s}\widetilde{\mathcal{L}}^n(a)\rho^s b)
=\tr(\rho^{1-s}\widetilde{\mathcal{L}}^{n-1}(a)\rho^s \Ll(b))=\dots
=\tr(\rho^{1-s}a\rho^s \Ll^n(b)).
\]
Multiplying by $t^n/n!$ and summing on $n$, we obtain (\ref{s-DB}) because
$\T_t=\sum_{n\ge 0} t^n\Ll^n/n!$ and $\widetilde{\T}_t=\sum_{n\ge 0} t^n\widetilde{\Ll}^n/n!$.

Finally (\ref{eq-L-Ltilde}) and (\ref{eq-L-Ltilde2}) follow from (\ref{eq-s-db-generators}) by the 
same arguments leading to the identity (\ref{eq-Ttilde-T}) starting from 
(\ref{s-DB}).
\qed\smallskip\\

We now characterise QMSs with $s$-dual for $s=0$ which is still a QMS. 
To this end, we start recalling some basic ingredient of Tomita-Takesaki 
theory. 

Let $L^2(\h)$ be the space of Hilbert-Schmidt operators on $\h$, 
with scalar product given by $\langle x,y\rangle_{HS}=\tr(x^*y)$.
If we set $\Omega=\rho^{1/2}\in L^2(\h)$ and 
$\pi_\rho(a): L^2(\h)\rightarrow L^2(\h)$ the left multiplication 
by $a\in\B$, then we obtain a representation of ${\mathcal{B}}(\h)$ 
on $L^2(\h)$ such that $\Omega$ is a cyclic and separating vector, and 
$\tr(\rho a)=\langle\Omega,\pi_\rho(a)\Omega\rangle_{HS}$ for every $a\in\B$. 
Under these hypothesis, identifying ${\mathcal{B}}(\h)$ with 
$\pi_\rho({\mathcal{B}}(\h))$, the modular operator $\Delta$ 
(see section 2.5.2 of Ref.\cite{olabra}) is defined on the 
dense set ${\mathcal{B}}(\h)\rho^{1/2}$ by 
\[
\Delta a\rho^{1/2}=\rho a\rho^{-1}\rho^{1/2}=\rho a \rho^{-1/2},
\] 
whereas a calculation shows that the modular group 
$(\sigma_t)_{t\in\mathbb{R}}$ on ${\mathcal{B}}(\h)$ is given by 
\[
\sigma_t(a)=\rho^{it}a\rho^{-it}.
\]

We recall that an element $a$ in $\B$ is analytic for $(\sigma_t)_t$ if there exists a strip $$I_\lambda=\{z\in\mathbb{C}\,\mid\,\mid\!\Im z\!\mid<\lambda\}$$ and a function $f:I_\lambda\rightarrow\B$ such that:
\begin{enumerate}\item $f(t)=\sigma_t(a)$ for all $t\in\mathbb{R}$;
\item $I_\lambda\ni z\rightarrow\tr(\eta f(z))$ is analytic for all $\eta\in L^1(\h)$ or, equivalently, $I_\lambda\ni z\rightarrow\langle u,f(z)v\rangle$ is analytic for all $u,v\in\h$.
\end{enumerate} 
We denote by $\!\textit{\Large{a}}\, $ the set of all analytic elements for $(\sigma_t)_t$.

It is a well known fact (Proposition $5$ of \cite{benfatto}) that $\textit{\Large{a}}\, \rho^{1/2}$ is a core for $\Delta$ and $\sigma_z(a)=\rho^{iz} a \rho^{-iz}\in\B$ for all $a\in\textit{\Large{a}}\,$ and $z\in\mathbb{C}$.\\
In particular, the modular automorphism $\sigma_{-i}$
on ${\mathcal{B}}(\h)$ is defined by $\sigma_{-i}(a)=\rho a\rho^{-1}$ for all $a\in\anal$ and it satisfies the following property

\begin{lemma}\label{autpos} If $\sigma_{-i}(a)=\alpha a$, then we have  
$\sigma_{-i}(a^*)=\alpha^{-1}a^*$ and $\alpha=\tr(\rho aa^*)/\tr(\rho a^*a)$.
In particular, every eigenvalue of $\sigma_{-i}$ is strictly positive.
\end{lemma}

\dimo
Let $\sigma_{-i}(a)=\alpha a$; then $\alpha\not=0$ for $\sigma_{-i}$ is invertible, 
and 
\[
\sigma_{-i}(a^*)=\rho a^*\rho^{-1}=(\sigma_{-i}^{-1}(a))^*=(\alpha^{-1}a)^*=\overline{\alpha^{-1}}a^*.
\]
But $\tr(\rho aa^*)=\tr(\rho a\rho^{-1}\rho a^*)=\alpha\tr(a\rho
a^*)=\alpha\tr(\rho a^*a)$, so that we obtain $\alpha=\tr(\rho
aa^*)/\tr(\rho a^*a)$ positive. Therefore,
$\sigma_{-i}(a^*)=\alpha^{-1}a^*$.
\qed\smallskip\\

%Given $X\in\B$ and a linear operator $Y$ on $\h$, $X$ and $Y$ commute means that $XY\subseteq YX$, i.e. $XY(u)=YX(u)$ for all $u$ which belongs to domain $D(Y)$ of $Y$.
We say that a linear bounded operator $X$ on $\B$ commute with $\sigma_{z}$ for some $z\in\mathbb{C}$ if $X(\sigma_z(a))=\sigma_z(X(a))$ for all $a\in\anal$. %since the set of analytic elements for $(\sigma_t)_t$ is dense in $\B$ (see \cite{olabra}), it follows that the commutation between $X$ and $\sigma_{-i}$, implies 
%$X(\rho a \rho^{-1})=\rho X(a) \rho^{-1}$ for all $a$ satisfying $\rho a \rho^{-1}\in\B$.\\

%We say that a linear bounded operator $X$ on $\B$ commute with $\sigma_{-i}$ if $X(\rho a \rho^{-1})=\rho X(a) \rho^{-1}$ for all $a\in D(\sigma_{-i})$, i.e. for all $a\in\B$ such that $\rho a\rho^{-1}\in\B$.

\smallskip
We can now show the following characterisation of QMSs whose $0$-dual is still  
a semigroup of {\em positive} linear maps, i.e. a QMS, adapting an argument 
from Majewski and Streater\cite{maje} (proof of Theorem 6). 

\begin{theorem}\label{th-dual-QMS-impl-commautmod} 
The following conditions are equivalent:
\begin{enumerate}
\item $\widetilde{\T}$ is a QMS;
\item any $\T_t$ commutes with $\sigma_{-i}$;
\item $\Ll$ commutes with $\sigma_{-i}$.
\end{enumerate}
If the above conditions hold, also the maps $\T_r$, ${\T}_{*r}$, $\tT_r$, $\tT_{*r}$ and the generators $\Ll$, $\Ll_*$, $\widetilde{\Ll}$,
$\widetilde{\Ll}_{*}$ commute with the homorphisms $\sigma_t$ for all $t,r\geq 0$.
\end{theorem} 
\dimo $(1)\Rightarrow (3)$ 
If $\widetilde{\T}$ is a QMS, then, in particular, $\widetilde{\T}_r$
satisfies $\widetilde{\T}_r(a^*)=\widetilde{\T}_r(a)^*$ for all $a\in{\mathcal{B}}(\h)$; 
therefore, by (\ref{eq-L-Ltilde2}) with $s=0$ and the same formula 
taking the adjoint we have
\[
\Ll(a)\rho=\widetilde{\Ll}_{*}(a\rho)\ \ \
\rho\Ll(a)=\widetilde{\Ll}_{*}(\rho a),
\] 
so that, replacing $a$ by $\rho a \rho^{-1}$,
\[\Ll(\rho a \rho^{-1})
=\widetilde{\Ll}_{*}((\rho a\rho^{-1})\rho)\rho^{-1}
=\rho\Ll_r(a)\rho^{-1}
\] 
for all $a\in\anal$. This means $\Ll\circ\sigma_{-i}=\sigma_{-i}\circ\Ll$ in the previous sense.

$(3)\Rightarrow (2)$ By induction we can show that $\Ll^n$ and $\sigma_{-i}$
commute for every $n\geq 0$; then, also $\T_t$ commute with $\sigma_{-i}$, 
for $\T_t=\exp({t\Ll})=\sum_{n\ge 0}t^n{\mathcal{L}}^n/n!$.

$(2)\Rightarrow (1)$ Let us define a contraction semigroup 
$(\widehat{T}_r)_{r\ge 0}$ on $L^2(\h)$ by 
\[
\widehat{T}_r(a\rho^{1/2})=\T_r(a)\rho^{1/2} 
\] 
for all $a\in{\mathcal{B}}(\h)$ and $r\ge 0$. Indeed, since 
$\T_r(a^*)\T_r(a)\le \T_r(a^*a)$ by the $2$-positivity of $\T_r$, we have  
\[
\Vert \widehat{T}_r(a\rho^{1/2})\Vert_{HS}^2 
=\tr\left(\rho^{1/2}\T_r(a^*)\T_r(a)\rho^{1/2}\right)
\le  \tr\left(\rho^{1/2}\T_r(a^* a)\rho^{1/2}\right)=\Vert a\rho^{1/2}\Vert_{HS}^2
\]
by the invariance of $\rho$ and the semigroup property follows from a straightforward 
algebraic computation. Condition $(2)$ implies then 
\begin{eqnarray*}
\widehat{T}_r(\Delta a\rho^{1/2})
&=&\widehat{T}_r(\rho a\rho^{-1}\rho^{1/2})
=\T_r(\rho a\rho^{-1})\rho^{1/2}=\rho\T_r(a)\rho^{-1}\rho^{1/2}\\ 
&=&\Delta\T_r(a)\rho^{1/2}=\Delta\widehat{T}_r(a\rho^{1/2})
\end{eqnarray*} 
for all $a\in\anal$, i.e. any map $\widehat{T}_r$ commutes with $\Delta$ ($\anal\rho^{1/2}$ is a core for $\Delta$). Therefore, by spectral calculus, 
$\widehat{T}_r$ also commutes with $\Delta^{it}$ for all $t\in\mathbb{R}$. It follows  
that $\T_r$ commutes with $\sigma_t$ for all $t\ge 0$. Thus 
\begin{eqnarray*}
\tr(\sigma_t(\T_{*r}(b))a)&=&\tr(\T_{*r}(b)\sigma_{-t}(a))=\tr(b\T_r(\sigma_{-t}(a)))\\ 
&=&\tr(b\sigma_{-t}(\T_r(a)))=\tr(\sigma_{t}(b)\T_r(a))\\ &=&\tr(\T_{*r}(\sigma_{t}(b))a)
\end{eqnarray*} 
for all $a$, $b$, i.e. also $\T_{*r}$ commutes with $\sigma_t$. Then, for all 
$r\ge 0$ and $t\in\mathbb{R}$, we get 
\begin{equation}
\label{T*sigma}\rho^{it}\T_{*r}(b)\rho^{-it}=\sigma_t(\T_{*r}(b))=\T_{*r}(\sigma_{t}(b))
=\T_{*r}(\rho^{it}b\rho^{-it}).
\end{equation}
We want to show that this equation holds for $b=\rho^{1/2}a\rho^{1/2}$ and for certain complex $t$. 
Since the maps 
\[
z \to\rho^{iz}=\hbox{\rm e}^{iz\ln\rho},\qquad z\to\rho^{-iz}
=\hbox{\rm e}^{-iz\ln\rho}
\] 
are analytic on $\Im z\leq 0$ and $\Im z\geq 0$ respectively, and the operator
\[
\rho^{i(t+is)}\rho^{1/2}a\rho^{1/2}\rho^{-i(t+is)}
=\rho^{it}\rho^{-s+1/2}a\rho^{1/2+s}\rho^{-it}
\] 
is trace class into the strip $1/2\leq s\leq 1/2$, both sides of equation (\ref{T*sigma}) have an analytic continuation 
into this strip, so that $\rho^{iz}\T_{*r}(b)\rho^{-iz}=\T_{*r}(\rho^{iz}b\rho^{-iz})$ 
%holds for all complex $z$ with $\mid\,\Im z\,\mid\leq 1/2$ and $b=\rho^{1/2}a\rho^{1/2}$. 
%Since the operator
%\[
%\rho^{i(t+is)}\rho^{1/2}a\rho^{1/2}\rho^{-i(t+is)}
%=\rho^{it}\rho^{-s+1/2}a\rho^{1/2+s}\rho^{-it}
%\] 
%is trace class into the strip $1/2\leq s\leq 1/2$, $\rho^{1/2}a\rho^{1/2}$ is an analytic element in this strip, and such a set is invariant under $\mathcal{T}_*$ for this commutes with $\sigma_t$. Therefore, both sides of equation (\ref{T*sigma}) have an analytic continuation 
%into the strip, so that $\rho^{iz}\T_{*r}(b)\rho^{-iz}=\T_{*r}(\rho^{iz}b\rho^{-iz})$ 
holds for all complex $z$ with $\mid\!\Im z\!\mid\leq 1/2$ and $b=\rho^{1/2}a\rho^{1/2}$. 
Taking $z=-i/2$, we get 
\begin{equation}
\rho^{1/2}\T_{*r}(\rho^{1/2}a\rho^{1/2})\rho^{-1/2}
=\T_{*r}(\rho^{1/2}\rho^{1/2}a\rho^{1/2}\rho^{-1/2})
=\T_{*r}(\rho a)=\rho\widetilde{\T}_r(a).
\end{equation} 
Hence 
\[
\widetilde{\T}_r(a)=\rho^{-1/2}\T_{*r}(\rho^{1/2}a\rho^{1/2})\rho^{-1/2},
\]
therefore any operator $\widetilde{\T}_r$ is completely positive and 
$\widetilde{\T}$ is a QMS.

The above arguments also prove the claimed commutation of semigroups, their generators 
and the homomorphisms $\sigma_t$.
\qed

\section{The generator of the dual semigroup}\label{sect-generator-quantum-dual}

Suppose now that the dual semigroup $\widetilde{\T}$ (for $s=0$) is a QMS with 
generator $\widetilde{\Ll}$. In this section we find the relationship between the 
operators $H, L_k$ and $\widetilde{G},\widetilde{L}_k$ which appear in the Lindblad 
representation of $\Ll$ and $\widetilde{\Ll}$. 

To this end, we start recalling the following result from 
Parthasarathy\cite{part} (Th. 30.16) on the representation of the generator of a uniformly 
continuous QMS in a special form of GKLS (Gorini, Kossakowski, Sudarshan, Lindblad) type.

\begin{theorem}\label{th-special-GKSL} 
Let $\Ll$ be the generator of a uniformly continuous QMS on ${\mathcal{B}}(\h)$ 
and let $\rho$ be any normal state on ${\mathcal{B}}(\h)$. Then there exist a bounded 
selfadjoint operator $H$ and a sequence $(L_k)_{k\geq 1}$ of elements in ${\mathcal{B}}(\h)$ 
such that: 
\begin{enumerate}
\item $\hbox{\em tr}(\rho L_k)=0$ for each $k\geq 1$,
\item $\sum_{k\geq 1}L^*_kL_k$ is strongly convergent, 
\item if $\sum_{k\geq 0}\vert c_k\vert^2<\infty$ and $c_0+\sum_{k\geq 1}c_kL_k=0$ 
for scalars $(c_k)_{k\geq 0}$ then $c_k=0$ for every $k\geq 0$,
\item $\mathcal{L}(a)=i[H,a]
      -\frac{1}{2}\sum_{k\geq 1}\left(L^*_kL_ka-2L^*_kaL_k+aL^*_kL_k\right)$ 
for all $a\in{\mathcal{B}}(\h)$.
\end{enumerate} 
Moreover, if $H^\prime,(L^\prime_k)_{k\ge 1}$ is another family of 
bounded operators in ${\mathcal{B}}(\h)$ with $H^\prime$
selfadjoint, then it satisfies conditions (1)-(4) if and only if the
lengths of sequences $(L_k)_{k\ge 1},(L^\prime_k)_{k\ge 1}$ are equal and
\[
H^\prime=H+\alpha,\qquad L^\prime_k=\sum_{j}u_{kj}L_j
\]
for some scalar $\alpha$ and a unitary matrix $(u_{kj})_{kj}$.
\end{theorem}

In our framework $\rho$ will always be a faithful normal $\T$-invariant state.

We now introduce a terminology in order to distinguish GKSL representations 
with properties (1) and (3) in Theorem \ref{th-special-GKSL}  from 
standard GKSL representations.

\begin{definition}\label{def-special-GKSL}
We call {\em special GKSL representation} with respect to a state $\rho$ 
by means of the operators $H,L_k$ any representation of $\Ll$ satisfying 
conditions (1),$\dots$,(4) of Theorem \ref{th-special-GKSL}. 
\end{definition}

\begin{remark}\rm Condition 3 of Theorem \ref{th-special-GKSL} means that 
$\{\unit, L_1,L_2,\ldots\}$ is a set of linearly independent elements of 
${\mathcal{B}}(\h)$. If dim$\h=d$, then the length of $(L_k)_{k\ge 1}$ in 
a special GKSL representation of $\Ll$ is at most $d^2-1$.
\end{remark}

We recall that we can also write $\Ll$ as
\[
\Ll(a)=G^*a+aG+\sum_kL^*_kaL_k,
\] 
where $G$ is the bounded operator on $\h$ defined by 
\begin{equation}\label{defG}
G=-iH-\frac{1}{2}\sum_kL^*_kL_k.
\end{equation} 

\begin{remark}\rm\label{rem-G-unique}
The last statement of Theorem \ref{th-special-GKSL} implies 
that, in a special GKSL representation of $\Ll$, the above operator $G$ is 
unique up to a purely imaginary multiple of the identity operator. Indeed 
the operator $G^\prime$ defined as in (\ref{defG}) replacing $H,L_k$ by 
$H^\prime, L^\prime_k$ satisfies
\begin{eqnarray*}
G^\prime 
& = & -iH -i\alpha -\frac{1}{2}\sum_{k,j,m}\bar{u}_{kj} u_{km} L^{*}_j L_m \\
& = & -iH -i\alpha -\frac{1}{2}\sum_{j,m}\left(\sum_k \bar{u}_{kj}u_{km}\right)L^{*}_j L_m \\
& = & -iH -i\alpha -\frac{1}{2}\sum_{j}L^{*}_j L_j = G - i\alpha 
\end{eqnarray*}
because the matrix $(u_{kj})_{kj}$ is unitary.
\end{remark}

Let $\mathsf{k}$ be a Hilbert space with Hilbertian dimension equal to the 
length of the sequence $(L_k)_k$ and let $(f_k)$ be an orthonormal basis of 
$\mathsf{k}$. Defining a linear bounded operator $L:\h\rightarrow\h\otimes \mathsf{k}$ 
by $Lu=\sum_kL_ku\otimes f_k$, Theorem \ref{th-special-GKSL} takes the following 
form (Theorem 30.12 Ref.\cite{part})

\begin{theorem}\label{DLMconL} 
If $\Ll$ is the generator of a uniformly continuous QMS on ${\mathcal{B}}(\h)$, 
then there exist an Hilbert space $\mathsf{k}$, a bounded linear operator 
$L:\h\rightarrow\h\otimes \mathsf{k}$ and a bounded selfadjoint operator $H$ 
in $\h$ satisfying the following:
\begin{enumerate}
\item $\Ll(x)=i[H,x]-\frac{1}{2}\left(L^*Lx
        -2L^*(x\otimes{\hbox{\em 1\kern-2.8truept l}}_\mathsf{k})L+xL^*L\right)$ 
         for all $x\in{\mathcal{B}}(\h)$;
\item the set 
      $\{(x\otimes{\hbox{\em 1\kern-2.8truept l}}_\mathsf{k})Lu:x\in{\mathcal{B}}(\h),\ u\in\h\}$ 
      is total in $\h\otimes \mathsf{k}$.
\end{enumerate}
\end{theorem}
\dimo Letting $Lu=\sum_kL_ku\otimes f_k$, where $(f_k)$ is 
an orthonormal basis of $\mathsf{k}$ and the $L_k$ are as in Theorem \ref{th-special-GKSL}, 
a simple calculation shows that condition (1) is fulfilled.

Suppose that there exists a non-zero vector
$\xi\in\{(x\otimes\unit_\mathsf{k})Lu:x\in{\mathcal{B}}(\h),\ u\in\h\}^\perp$; 
then $\xi=\sum_kv_k\otimes f_k$ with $v_k\in\h$ and 
\[
0=\langle \xi,(x\otimes\unit_\mathsf{k})Lu\rangle=\sum_k\langle v_k,xL_ku\rangle
 =\sum_k\langle L^*_kx^*v_k,u\rangle
\] 
for all $x\in{\mathcal{B}}(\h)$, $u\in\h$. Hence, $\sum_kL^*_kx^*v_k=0$.
Since $\xi\not=0$, we can suppose $\Vert v_1\Vert=1$; then, putting $p=|{v_1}\rangle\langle{v_1}|$ 
and $x=py^*$, $y\in{\mathcal{B}}(\h)$, we get 
\begin{equation}\label{noli}
0=L^*_1yv_1 +\sum_{k\geq 2}\langle v_1,v_k\rangle L^*_kyv_1
=\bigg(L^*_1
+\sum_{k\geq 2}\langle v_1,v_k\rangle L^*_k\bigg)yv_1.
\end{equation} 
Since $y\in{\mathcal{B}}(\h)$ is arbitrary, equation (\ref{noli}) contradicts the 
linear independence of the $L_k$'s. Therefore the set in (2) must be total. 
\qed\smallskip\\ 

We now study the generator $\Ll$ of QMS ${\T}$ whose dual $\widetilde{\T}$ 
is a QMS. As a first step we find an explicit form for the operator $G$ 
defined by (\ref{defG}).

\begin{proposition}\label{carattG} 
If $\Ll(a)=G^*a+aG+\sum_jL^*_jaL_j$ is a special GKSL representation of $\Ll$ 
and $\rho$ is the $\T$-invariant state (\ref{eq-rho}) then 
\begin{eqnarray}
G^*u &=&\sum_{k\geq 1}\rho_k\Ll(\ketbra{u}{e_k})e_k-\hbox{\em tr}(\rho G)u \label{eq-G-star-u}\\
 Gv  &=&\sum_{k\geq 1}\rho_k\Ll_*(\ketbra{v}{e_k})e_k-\hbox{\em tr}(\rho G^*)v\label{eq-G-v}
\end{eqnarray} 
for every $u,v\in\h$.
\end{proposition}
\dimo 
Since $\Ll(|{u}\rangle\langle{v}|) =|{G^*u}\rangle\langle{v}|
+|u\rangle\langle{Gv}|+\sum_j |{L^*_ju}\rangle\langle{L^*_jv}|$, 
letting $v=e_k$ we have $G^*u =|{G^*u}\rangle\langle{e_k}|e_k$ and
\[
G^*u =\Ll(|u\rangle\langle e_k|)e_k  
- \sum_{j}\langle e_k,L_je_k\rangle L^*_ju 
- \langle  e_k,Ge_k\rangle u.
\] 
Multiplying both sides by $\rho_k$ and summing on $k$, we find then
\begin{eqnarray*}
G^*u&=&\sum_{k\geq 1}\rho_k\Ll(|u\rangle\langle e_k|)e_k
 - \sum_{j,k}\rho_k\langle e_k,L_je_k\rangle L^*_ju-\sum_{k\geq 1}\rho_k\langle  e_k,Ge_k\rangle u\\
&=&\sum_{k\geq 1}\rho_k\Ll(\ketbra{u}{e_k})e_k-\sum_j\tr(\rho L_j)L^*_ju-\tr(\rho G)u
\end{eqnarray*} 
and (\ref{eq-G-star-u}) follows since $\tr(\rho L_j)=0$.  Computing the adjoint of $G$ we 
find immediately (\ref{eq-G-v}).
\qed

\begin{proposition}\label{Gcomm} 
Let $\widetilde{\T}$ be the $s$-dual of a QMS $\T$ with generator $\widetilde{\Ll}$.
If $G$ and $\widetilde{G}$ are the operators (\ref{eq-G-v}) in some special GKSL 
representations of $\Ll$ and $\widetilde{\Ll}$ then
\begin{equation}\label{reltildeG}
\widetilde{G}\rho^{s}=\rho^sG^*+\left(\hbox{\em tr}(\rho G)-\hbox{\em tr}(\rho\widetilde{G}^*)\right)\rho^{s}.
\end{equation} 
Moreover, we have 
$\hbox{\em tr}(\rho G)-\hbox{\em tr}(\rho\widetilde{G}^*)=ic_\rho$ for 
some $c_\rho\in\mathbb{R}$.
\end{proposition}

\dimo 
The identities (\ref{eq-G-v}) and (\ref{eq-L-Ltilde})  yield
\begin{eqnarray*} 
\widetilde{G}\rho^{s}
 &=&\sum_{k\geq 1}\widetilde{\Ll}_*(\rho^{s}\ketbra{v}{\rho_k^{1-s}e_k})\rho_k^s e_k-\tr(\rho \widetilde{G}^*)\rho^{s}v\\
 &=&\sum_{k\geq 1}\widetilde{\Ll}_*(\rho^s(\ketbra{v}{e_k})\rho^{1-s})\rho^se_k-\tr(\rho \widetilde{G}^*)\rho^{s}v\\
 &=&\sum_{k\geq 1}\rho^s\Ll(\ketbra{v}{e_k})\rho^{1-s}\rho^se_k-\tr(\rho \widetilde{G}^*)\rho^{s}v\\
 &=&\rho^sG^*v+\left(\tr(\rho G)-\tr(\rho \widetilde{G}^*)\right)\rho^{s}v.
\end{eqnarray*} Therefore, we obtain (\ref{reltildeG}). 

Right multiplying equation (\ref{reltildeG}) by $\rho^{1-s}$ we have
$\widetilde{G}\rho=\rho^sG^*\rho^{1-s}+\left(\tr(\rho G)-\tr(\rho\widetilde{G}^*)\right)\rho$, 
so that taking the trace,
\begin{eqnarray*}
\tr(\rho G)-\tr(\rho\widetilde{G}^*)
&=&\tr(\widetilde{G}\rho)-\tr(\rho^sG^*\rho^{1-s})\\
&=&\tr(\widetilde{G}\rho)-\tr(G^*\rho)=-\overline{\left(\tr(\rho G)-\tr(\rho\widetilde{G}^*)\right)};
\end{eqnarray*} 
this proves that $\tr(\rho G)-\tr(\rho\widetilde{G}^*)=ic_\rho$ for some real constant $c_\rho$.
\qed

\begin{proposition}\label{prop-G-comm-rho}
Let $\widetilde{\T}$ be the $0$-dual of a QMS $\T$ and let 
\[
\Ll(a)=G^*a+\sum_jL^*_jaL_j+aG, \qquad\ 
\widetilde{\Ll}(a)=\widetilde{G}^*a+\sum_j\widetilde{L}^*_ja\widetilde{L}_j+a\widetilde{G}
\] 
be special GKSL representations of $\Ll$ and $\widetilde{\Ll}$. 
Then:
\begin{enumerate}
\item $\widetilde{G}=G^*+ic$ with $c\in\mathbb{R}$,
\item both $G$ and $\widetilde{G}$ commute with $\rho$,
\item $\sum_k L_k^*L_k$, $\sum_k \widetilde{L}^*_k\widetilde{L}_k$, $H$ and $\widetilde{H}$ 
commute with $\rho$.
\end{enumerate}
\end{proposition}
\dimo 
$(1)$ It follows by Proposition \ref{Gcomm} for $s=0$ and Theorem \ref{th-special-GKSL}, 
Remark \ref{rem-G-unique}. Indeed, in any special GKSL representations of $\Ll$ 
and $\widetilde{\Ll}$, $G$ and $\widetilde{G}$ are unique up to a purely imaginary 
multiple of the identity operator.

$(2)$ Let $G$ and $\widetilde{G}$ be the operators (\ref{eq-G-v}) and in the given special 
GKSL representations of $\Ll$ and $\widetilde{\Ll}$. Since $\widetilde{\Ll}_*(\rho a)=\rho\Ll(a)$ 
holds (for $\widetilde{\T}$ is a QMS), we have 
\begin{eqnarray*}
\widetilde{G}\rho v 
& = & \sum_{k=1}^n 
\rho_k\widetilde{\Ll}_*(\rho|{v}\rangle\langle{\rho e_k}|)e_k-\tr(\rho \widetilde{G}^*)\rho \\
& = &\sum_{k=1}^n\rho_k\rho\Ll(|{v}\rangle\langle{e_k}|)e_k-\tr(\rho \widetilde{G}^*)\rho v \\
& = &\rho\left(G^*v+\tr(\rho G)v\right)-\tr(\rho \widetilde{G}^*)\rho v
\end{eqnarray*} 
for every $v\in\h$, that is $\widetilde{G}\rho=\rho G^*+ic_\rho\rho$. But
$G$ and $\widetilde{G}$ are the operators (\ref{eq-G-v}), therefore by (1) 
we have also $\widetilde{G}\rho=G^*\rho+ic_\rho\rho$, and so $G^*\rho=\rho G^*$. 
This, together with Remark \ref{rem-G-unique}, clearly implies $(2)$.

$(3)$ Follows from $(2)$ by decomposing $G$ and $\widetilde{G}$ into their self-adjoint 
and anti self-adjoint parts. 
\qed\medskip\\

We now study the properties of the $L_k$ when $\widetilde{\T}$ is a QMS.

\begin{lemma}\label{Lautovettfinitodim}
With the notations of Theorem \ref{DLMconL}, if $\h$ is finite-dimensional %Hilbert space, $L:\h\rightarrow\h\otimes\mathsf{k}}$, $Lu=\su and $\rho$ a faithful state on $\h$ such that the set $\{(x\otimes\unit_{\mathsf{k}})Lu\,\mid\, x\in{\mathcal{B}}(\h),\ u\in\h\}$ is total in $\h\otimes\mathsf{k}$. 
then the equation 
\begin{equation}\label{Lsigma-i}
\sum_kL^*_kaL_k=\sum_k\rho^{-1}L^*_k\rho a\rho^{-1}L_k\rho\ \ \forall\ a\in\B
\end{equation}
implies $\rho L_k\rho^{-1}=\lambda_kL_k$ and $\rho L^*_k\rho^{-1}=\lambda^{-1}_kL^*_k$ for some positive $\lambda_k$.
\end{lemma}
\dimo
Define two linear maps $X_1,X_2$ on $\h\otimes\mathsf{k}$ by
\begin{eqnarray*}
X_1(x\otimes\unit_\mathsf{k})Lu 
& = &(x\otimes\unit_\mathsf{k})(\rho\otimes\unit_\mathsf{k})L\rho^{-1}u \\
X_2(x\otimes\unit_\mathsf{k})Lu 
& = & (x\otimes\unit_\mathsf{k})(\rho^{-1}\otimes\unit_\mathsf{k})L\rho u
\end{eqnarray*}
for all $x\in{\mathcal{B}}(\h)$ and $u\in\h$. %where $L:\h\rightarrow\h\otimes \mathsf{k}$, 
%$Lu=\sum_kL_ku\otimes f_k$, $(f_k)_k$ o.n.b. of $\mathsf{k}$, are defined as 
%in Theorem \ref{DLMconL}.

We postpone to Lemma \ref{Xbendef} the proof that $X_1$ and $X_2$ are well defined 
on the total (Theorem \ref{DLMconL}) set $\{(x\otimes\unit_\mathsf{k})Lu\,\mid\,
x\in{\mathcal{B}}(\h),\ u\in\h\}$ in $\h\otimes\mathsf{k}$. We can now extend $X_1$ 
and $X_2$ to a bounded operator on $\h\otimes \mathsf{k}$. Moreover, 
$(\ref{eq-cp-part-comm-with-mod-aut})$ implies
\begin{eqnarray*}
\langle X_1(x\otimes\unit_\mathsf{k})L u,X_2(y\otimes\unit_\mathsf{k})L v\rangle
&=&\sum_k\langle u,\rho^{-1} L^*_k\rho x^*y\rho^{-1}L_k\rho v\rangle\\
&=&\sum_k\langle u,L^*_kx^*yL_kv\rangle\\ 
&=&\langle(x\otimes\unit_\mathsf{k})Lu,(y\otimes\unit_\mathsf{k})Lv\rangle
\end{eqnarray*}
for all $x,y\in{\mathcal{B}}(\h)$ and $u\in\h$. As a consequence we have 
$X_1^*X_2=\unit_{\h\otimes\mathsf{k}}$. 

By the definition of $X_1,X_2$ we have also $X_j(y\otimes\unit_k)
=(y\otimes\unit_k)X_j$ for $j=1,2$.
Therefore $X_j$ can be written in the form $\unit_{\h}\otimes Y_j$ for some
invertible bounded map $Y_j$ on $\mathsf{k}$ satisfying $Y_1^*Y_2=\unit_\mathsf{k}$. 

The definition of $X_1,X_2$ implies then
\begin{equation}\label{eq-Y-and-L}
(\unit_{\h}\otimes Y_1)L = (\rho\otimes\unit_\mathsf{k})L\rho^{-1}, \qquad 
\left(\unit_{\h}\otimes Y_1^*\right)^{-1}L=(\rho^{-1}\otimes\unit_\mathsf{k})L\rho,
\end{equation}
right multiplying by $\rho$ and left multiplying by 
$(\rho\otimes\unit_\mathsf{k})$ the first and the second identity we find
\[
(\unit_{\h}\otimes Y_1)L\rho = (\rho\otimes\unit_\mathsf{k})L, \qquad 
\left(\unit_{\h}\otimes Y_1^*\right)^{-1}(\rho\otimes\unit_\mathsf{k})L=L\rho.
\]
Writing the second as 
$(\rho\otimes\unit_\mathsf{k})L=\left(\unit_{\h}\otimes Y_1^*\right)L\rho$
we obtain
\[
(\unit_{\h}\otimes Y_1)L\rho = (\rho\otimes\unit_\mathsf{k})L
=\left(\unit_{\h}\otimes Y_1^*\right)L\rho.
\]
Since $\rho$ is faithful, it follows that $(\unit_{\h}\otimes Y_1)L 
=\left(\unit_{\h}\otimes Y_1^*\right)L$ 
(and also $(\unit_{\h}\otimes Y_1)(x\otimes\unit_{\mathsf{k}})L
=\left(\unit_{\h}\otimes Y_1^*\right)(x\otimes\unit_{\mathsf{k}})L$
for all $x$) proving that $Y_1$ is self-adjoint.

Therefore, there exist non-zero $\lambda_k\in\mathbb{R}$ and a 
unitary operator $U$ on $\mathsf{k}$ such that $Y_1=U^*DU$ with
$D=\mbox{diag}(\lambda_1,\lambda_2,\dots)$. The identities 
(\ref{eq-Y-and-L}) yield then
\begin{eqnarray*}
(\unit_\h\otimes DU)L
& = &(\unit_\h\otimes U)(\rho\otimes\unit_\mathsf{k})L\rho^{-1} 
 =(\rho\otimes \unit_\mathsf{k})(\unit_\h\otimes U)L\rho^{-1}\\
(\unit_\h\otimes D^{-1}U)L 
& = & (\unit_\h\otimes U)(\rho^{-1}\otimes \unit_\mathsf{k})L\rho
= (\rho^{-1}\otimes \unit_\mathsf{k})(\unit_\h\otimes U)L\rho
\end{eqnarray*}

Thus, putting $L^\prime=UL$, or, more precisely $L^\prime_k=\sum_\ell u_{k\ell}L_\ell$
for all $k$, we have
\[
\rho L^\prime_k\rho^{-1}=\lambda_kL^\prime_k\ \ \mbox{and}\ \
\rho^{-1}L^\prime_k\rho=\lambda_k^{-1}L^\prime_k
\] 
for every $k$. To conclude the proof it suffices to recall that $\lambda_k>0$ 
by Lemma \ref{autpos}, since the above identities mean that 
$\lambda_k$ is an eigenvalue of $\sigma_{-i}$. 
\qed\medskip\\ 

We now check that the maps $X_1,X_2$ introduced in the proof of Lemma \ref{Lautovettfinitodim}
are well defined. 
\begin{lemma}\label{Xbendef} 
%Suppose that $\Ll$ and $\sigma_{-i}$ commute and let 
%$x_1,\ldots,x_m\in{\mathcal{B}}(\h)$, $u_1,\ldots,u_m\in\h$.
%With the notations of Proposition \ref{Lautovett}, if  
With the notations of Lemma \ref{Lautovettfinitodim}, if $\h$ is finite-dimensional and equation $(\ref{Lsigma-i})$ holds, then 
\begin{equation}\label{eq-x-L-zero}
\sum_{j=1}^m(x_j\otimes{\hbox{\em 1\kern-2.8truept l}}_\mathsf{k})Lu_j=0
\end{equation} 
for $x_1,\ldots,x_m\in{\mathcal{B}}(\h)$, $u_1,\ldots,u_m\in\h$ implies:
\begin{enumerate}
\item $\sum_{j=1}^m(x_j\otimes{\hbox{\em 1\kern-2.8truept l}}_\mathsf{k})
(\rho\otimes{\hbox{\em 1\kern-2.8truept l}}_\mathsf{k})L\rho^{-1}u_j=0$;
\item $\sum_{j=1}^m(x_j\otimes{\hbox{\em 1\kern-2.8truept l}}_\mathsf{k})
(\rho^{-1}\otimes{\hbox{\em 1\kern-2.8truept l}}_\mathsf{k})L\rho u_j=0$.
\end{enumerate}
\end{lemma}
\dimo 
Suppose that (\ref{eq-x-L-zero}) holds.  Taking the adjoint of 
(\ref{eq-cp-part-comm-with-mod-aut}) we find
\[
\sum_k\rho L^*_k\rho^{-1}a\rho L_k\rho^{-1}=\sum_kL^*_kaL_k
\] 
for every $a\in{\mathcal{B}}(\h)$ and compute
\begin{eqnarray*}
& & \langle(y\rho^{-1}\otimes\unit_\mathsf{k})L\rho v,
     \sum_{j=1}^m(x_j\otimes\unit_\mathsf{k})(\rho\otimes\unit_\mathsf{k})L\rho^{-1}u_j\rangle \\
& & =\sum_{j=1}^m\sum_k\langle v,L^*_ky^*x_jL_ku_j\rangle
    =\langle(y\otimes\unit_\mathsf{k})Lv,\sum_{j=1}^m(x_j\otimes\unit_\mathsf{k})Lu_j\rangle=0
\end{eqnarray*}
for all $y\in{\mathcal{B}}(\h)$ and $v\in\h$. But the set 
$S=\{(y\rho^{-1}\otimes\unit_\mathsf{k})L\rho v\,\mid\, y\in{\mathcal{B}}(\h),\ v\in\h\}$ 
is total in $\h\otimes \mathsf{k}$, because 
$\{(y\otimes\unit_\mathsf{k})Lv\,\mid\, y\in{\mathcal{B}}(\h),\ v\in\h\}$ 
is total (Theorem \ref{DLMconL}) and the maps $y\mapsto y\rho^{-1}$, $v\mapsto \rho v$ 
are bijective. This proves (1). The proof of (2) is similar and we omit it.
\qed

\begin{proposition}\label{Lautovett} 
Suppose that $\Ll$ and $\sigma_{-i}$ commute. 
Then there exists a special GKSL representation of $\Ll$ in which, for all k, 
we have
\[
\rho L_k=\lambda_k L_k\rho,\quad \rho L^*_k=\lambda^{-1}_k L^*_k\rho,
\qquad \lambda_k>0.
\] 
\end{proposition}
\dimo %Define $p_n:=\sum_{k\leq n}\ee{e_k}{e_k}$ for $n\geq 0$ and consider $p_n\rho^{-1}$; since $$p_n\rho^{-1}u=\rho^{-1}p_nu=\sum_{k\leq n}\rho_k^{-1}\ee{e_k}{e_k}$$ for all $u\in D(\rho^{-1})$ and $\rho^{-1}p_n\in\B$, then we can extend $p_n\rho^{-1}$ to a continue operator on $\h$. It follows that $\rho p_nap_n\rho^{-1}$ belongs to $\B$ for every $a\in\B$, i.e $b:=p_nap_n\in D(\sigma_{-i})$.\\
Define $p_n:=\sum_{k\leq n}\ee{e_k}{e_k}$ for $n\geq 0$ and consider $u,v\in\h$, $a\in\B$; since the map 
$$
z\rightarrow\langle u,\rho^{iz}p_nap_n\rho^{-iz}\rangle=\sum_{k,h\leq n}\rho_k^{iz}\rho_h^{-iz}\langle u,e_k\rangle\langle e_k,ae_h\rangle\langle e_h,v\rangle
$$
is analytic on $\mathbb{C}$, then $b:=p_nap_n$ belongs to $\anal$ for all $n\geq 0$.
As a consequence, since $\Ll$ and $\sigma_{-i}$ commute, we have 
$\Ll( b)=\rho^{-1}\Ll(\rho b\rho^{-1})\rho$, so that 
\begin{eqnarray*}
& & i[H,b]-\frac{1}{2}\sum_k\left(L^*_kL_kb-2L^*_kbL_k+bL^*_kL_k\right)
 = \rho^{-1}[H,\rho b\rho^{-1}]\rho \\
& & \qquad\qquad -\frac{1}{2}\sum_k\left(\rho^{-1}L^*_kL_k\rho b
-2\rho^{-1}L^*_k\rho b\rho^{-1}L_k\rho+b\rho^{-1}L^*_kL_k\rho\right).
\end{eqnarray*}
Both $H$ and $\sum_k L^*L_k$ commute with $\rho$ by Proposition \ref{prop-G-comm-rho} (3). 
We have then
\begin{equation}\label{eq-cp-part-comm-with-mod-aut}
\sum_kL^*_kp_nap_nL_k=\sum_k\rho^{-1}L^*_k\rho p_nap_n\rho^{-1}L_k\rho,
\end{equation} 
and so
$$\sum_kp_nL^*_kp_nap_nL_kp_n=\sum_kp_n\rho^{-1}L^*_k\rho p_nap_n\rho^{-1}L_k\rho p_n$$ 
for all $a\in\B$, $n\geq 0$, right and left multiplying (\ref{eq-cp-part-comm-with-mod-aut}) by $p_n$. Remembering that $p_n\rho^{-1}=\rho^{-1}p_n$ on $\rho(\h)$ and setting $L_{(n)k}:=p_nL_kp_n$, $\rho_{(n)}:=\rho p_n$, the above equality gives
$$\sum_k\rho_{(n)}^{-1}L^*_{(n)k}\rho_{(n)} b\rho_{(n)}^{-1}L_{(n)k}\rho_{(n)}=\sum_kL^{*}_{(n)k}bL_{(n)k}$$ for all $b\in\mathcal{B}(p_n(\h))$. 

But $\rho_{(n)}$ is faithful on the finite-dimensional Hilbert space $p_n(\h)$ and $\{(x\otimes\unit_\mathsf{k})L_{(n)}u\,\mid\,
x\in{\mathcal{B}}(p_n(\h)),\ u\in p_n(\h)\}$ is total in $p_n(\h)\otimes\mathsf{k}$, therefore Lemma \ref{Lautovettfinitodim} assures that
$$\rho_{(n)}L_{(n)k}\rho^{-1}_{(n)}=\lambda_{k,n}L_{(n)k}\quad\rho_{(n)}L^*_{(n)k}\rho^{-1}_{(n)}=\lambda^{-1}_{k,n}L^*_{(n)k}$$
for some $\lambda_{k,n}>0$, i.e.
\begin{equation}\label{autovconn}
\rho p_nL_kp_n=\lambda_{k,n}p_nL_k\rho p_n,\quad \rho p_n L^*_kp_n=\lambda^{-1}_{k,n}p_n L^*_k\rho p_n.
\end{equation}
Since $(p_n)_n$ is an increasing sequence of projections, this implies $\lambda_{k,n}=\lambda_k$ for $n>>0$, and then, letting $n\rightarrow\infty$ in equation (\ref{autovconn}), we obtain 
\[
\rho L_k=\lambda_k L_k\rho,\quad \rho L^*_k=\lambda^{-1}_k L^*_k\rho,
\] 
for $(p_n)_n$ converges to $\unit$ in the strong operator topology.
\qed

\begin{definition}\label{def-privileged-GKSL}
Let $\Ll$ be the generator of a QMS and let $\rho$ be a faithful normal state.
A special GKSL representation of $\Ll$ with respect to $\rho$ by means of operators 
$H$, $L_k$ is called {\em privileged} if their operators $L_k$ satisfy 
$\rho L_k=\lambda_kL_k\rho$ and $\rho L^*_k=\lambda^{-1}_kL^*_k\rho$ for 
some $\lambda_k>0$ and $H$ commutes with  $\rho$.
\end{definition}

\begin{remark}\rm The operator $\sum_{k}L^*_kL_k$ (the self-adjoint part of $G$) in 
a privileged GKSL representation clearly commutes with $\rho$. Moreover, the constants 
$\lambda_k$ are determined by the eigenvalues of  $\rho$. Indeed, writing $\rho$ as 
in (\ref{eq-rho}), the identity $\rho L_k=\lambda_k L_k\rho$ yields 
\[
\rho_j\langle e_j, L_k e_m\rangle 
=  \langle  e_j, \rho L_k e_m \rangle 
= \lambda_k\langle e_j, L_k\rho e_m \rangle=\lambda_k \rho_m\langle e_j, L_k e_m \rangle.
\]
Therefore $\lambda_k = \rho_j\rho_m^{-1}$ for all $j,m$ such that 
$\langle e_j, L_k e_m\rangle \not=0$.
In particular, if we write $\rho=\hbox{\rm e}^{-H_S}$ for some bounded selfadjoint operator 
$H_S=\sum_i \varepsilon_j|e_j\rangle\langle e_j|$ on $\h$, we find 
$\lambda_k=\hbox{\rm e}^{\varepsilon_m-\varepsilon_j}$.
\end{remark}

\begin{proposition}\label{prop-non-uniqueness-priv-GKSL}
Given two privileged GKSL of $\Ll$ with respect to the same state $\rho$ by means 
of operators $H,L_k$ and $H^\prime,L_k^\prime$, with $D=\hbox{\em diag}(\lambda_1,
\lambda_2,\dots)$ and $D^\prime=\hbox{\em diag}(\lambda_1^\prime,\lambda_2^\prime,\dots)$, 
there exists a unitary operator $V$ on $\mathsf{k}$ and $\alpha\in\mathbb{R}$ such that 
\[
H^\prime = H + \alpha, \qquad L^\prime = ({\hbox{\em 1\kern-2.8truept l}}_\h\otimes V)L, 
\quad D^\prime = VDV^*.
\]
\end{proposition}

\dimo
By Theorem \ref{th-special-GKSL} there exist $\alpha\in\mathbb{R}$ and a unitary $V$ 
on $\mathsf{k}$ such that $H^\prime=H+\alpha$ and $L^\prime=(\unit_\h\otimes V)L$. 
Since both families $H,L_k$ and $H^\prime,L_k^\prime$ give privileged GKSL representations 
with respect to the same state $\rho$, we have
\[
(\rho\otimes\unit_{\mathsf{k}})L= (\unit_\h\otimes D)L\rho,\qquad
(\rho\otimes\unit_{\mathsf{k}})L^\prime= (\unit_\h\otimes D^\prime)L^\prime\rho.
\]
Left multiplying the first identity by $(\unit_\h\otimes V)$ and replacing $L^\prime$ 
by $VL$ in the second we find
\[
(\rho\otimes\unit_{\mathsf{k}})(\unit_\h\otimes V)L= (\unit_\h\otimes VD)L\rho, \quad
(\rho\otimes\unit_{\mathsf{k}})(\unit_\h\otimes V)L= (\unit_\h\otimes D^\prime V)L\rho
\]
It follows that $VD = D^\prime V$, i.e. $D^\prime=VDV^*$.
\qed

\begin{remark}\rm
The identity $D^\prime=VDV^*$ means that $V$ is a change of coordinates that transforms 
$D$ into another diagonal matrix; in particular, if $D=\hbox{\em diag}(\lambda_1,
\lambda_2,\dots)$ and $D^\prime=\hbox{\em diag}(\lambda_1^\prime,\lambda_2^\prime,\ldots)$, we have $$\lambda^\prime_i\langle f_i,Vf_j\rangle=\lambda_j\langle f_i,Vf_j\rangle.$$ Since $V$ is a unitary operator this implies that, when the $\lambda_k$ are all different, for every $i$ there exists a unique $j$ such that $\langle f_i,Vf_j\rangle\not=0$ and for every $j$ there exists a unique $i$ such that $\langle f_i,Vf_j\rangle\not=0$. Thus $$Vf_j=\hbox{\rm e}^{i\theta_{\sigma(j)}}f_{\sigma(j)}\ \ \mbox{and}\ \ L^\prime_k=\hbox{\rm e}^{i\theta_{\sigma(k)}}L_{\sigma(k)}$$ with $\theta_{\sigma(j)}\in\mathbb{R}$ and $\sigma$ a permutation. 
Therefore, when the $\lambda_k$ are all different, 
privileged GKSL representations of $\Ll$, if exist, are unique up to a permutation of 
the operators $L_k$, a multiplication of each $L_k$ by a phase $\hbox{\rm e}^{i\theta_k}$ 
and a constant $\alpha$ in the Hamiltonian $H$.\\
If some $\lambda_k$'s 
are equal, then also unitary transformations of subspaces of $\mathsf{k}$ associated 
with the same $\lambda_k$'s are allowed.
\end{remark}

The results of this section are summarized by the following 

\begin{theorem}\label{caratttildeQMS} 
The $0$-dual semigroup $\widetilde{\T}$ of a QMS $\T$ generated by $\Ll$ with faithful 
normal invariant state $\rho$ is a QMS if and only there exists a privileged GKSL
representation of $\Ll$ with respect to $\rho$.
\end{theorem} 
\dimo If $\widetilde{\T}$ is a QMS, then $\Ll$ commutes with $\sigma_{-i}$ 
by Theorem \ref{th-dual-QMS-impl-commautmod} and so there exists a special GKSL 
representation of $\Ll$ by Propositions \ref{Lautovett} and \ref{prop-G-comm-rho}.

The converse is trivial.
\qed\smallskip\\

We now exhibit an example of semigroup whose dual is not a QMS.

\begin{example}\rm \label{tildenoQMS} 
Consider the semigroup $\T$ on $M_2(\mathbb{C})$ generated by 
\begin{eqnarray*}
\Ll(a)=i\,\frac{\Omega}{2}\,[\sigma_1,a]
-\frac{\mu^2}{2}\left(\sigma^+\sigma^-a-2\sigma^+a\sigma^-+a\sigma^+\sigma^-\right),
\end{eqnarray*} 
where $\mu>0$, $\Omega\in\mathbb{R}$, $\Omega\not=0$ and $\sigma_k$ are the Pauli matrices and  
$\sigma^{\pm}=\sigma_1 \pm i\sigma_2$ are the raising and lowering operator.

A straightforward computation shows the state 
\[
\rho=
\frac{1}{2}\left(\unit+\frac{2\mu^2\Omega}{2\Omega^2+\mu^4}\sigma_2
-\frac{\mu^4}{2\Omega^2+\mu^4}\sigma_3\right)
=\frac{1}{2\Omega^2+\mu^4}
\left(\begin{array}{cc}\Omega^2&-i\mu^2\Omega\\ i\mu^2\Omega& \Omega^2+\mu^4\end{array}\right)
\] 
is invariant and faithful. The generator $\Ll$ can be written in a special GKSL form 
(with respect to the invariant state $\rho$) with 
\[
L_1=\mu\sigma^{-}-\frac{\mu}{2}\,\tr(\rho\sigma^{-})\unit=\mu\sigma^{-}+i\,
\frac{\mu^3\Omega}{2\left(2\Omega^2+\mu^4\right)}\,,\qquad H=\left(\frac{\Omega}{2}+\frac{\mu^4\Omega}{2\Omega^2+\mu^4}\right)\sigma_1.
\]
The dual semigroup $\widetilde{\T}$ of $\T$ is not 
a QMS because $H$ does not commute with $\rho$.
\end{example}
We now establish the relationship between the privileged GKSL representations of 
a generator $\Ll$ and its $0$-dual $\widetilde{\Ll}$.

\begin{theorem}\label{HLtilde} 
If $\widetilde{\T}$ is a QMS, then, for every privileged GKSL representation of $\Ll$, 
by means of operators $H, L_k$, there exists a privileged GKSL representation of $\widetilde{\Ll}$,
by means of operators $\widetilde{H},\widetilde{L}_k$ such that:
\begin{enumerate} 
\item $\widetilde{H}=-H-\alpha$ for some $\alpha\in\mathbb{R}$;
\item $\widetilde{L}_k= \lambda^{-1/2}_k L^*_k$ for some $\lambda_k>0$.
\end{enumerate} 
\end{theorem} 

\dimo Consider a privileged GKSL representation of $\Ll$
\[
\Ll(a)=i[H,a]
-\frac{1}{2}\sum_{k\geq 1}\left(L^*_kL_ka-2L^*_kaL_k+aL^*_kL_k\right),
\] 
with $H\rho=\rho H$ and $\rho L_k\rho^{-1}=\lambda_kL_k$, 
$\rho L^*_k\rho^{-1}=\lambda^{-1}_kL^*_k$ for some $\lambda_k>0$.

Since $\rho\widetilde{\Ll}(a)=\Ll_*(\rho a)$, we have 
\begin{eqnarray*}
\rho\widetilde{\Ll}(a)
&=&-i[H,\rho a]
-\frac{1}{2}\sum_k\left(\rho aL^*_kL_k-2L_k\rho aL^*_k+L^*_kL_k\rho a\right)\\ 
&=& -i\rho[{H},a]
-\frac{1}{2}\sum_k\left(\rho aL^*_kL_k-2\lambda^{-1}_k\rho L_kaL^*_k+L^*_kL_k\rho a\right).
\end{eqnarray*} 
But $\rho$ is $\T$-invariant and commutes with $H$ thus 
$\sum_k\rho L^*_kL_k=\sum_kL^*_kL_k\rho=\sum_kL_k\rho L^*_k=\sum_k\lambda_k^{-1}\rho L_kL^*_k$. It follows that $\sum_kL^*_kL_k=\sum_k\lambda_k^{-1}L_kL^*_k$ and
\begin{eqnarray*}
\rho\widetilde{\Ll}(a)
&=&\rho\left(-i[{H},a]
-\frac{1}{2}\sum_k\left(a\lambda_k^{-1}L_k
L^*_k-2\lambda^{-1}_kL_kaL^*_k+\lambda_k^{-1}L_kL^*_k a\right)\right).
\end{eqnarray*} 
Therefore, putting $\widetilde{H}=-H-\alpha$ ($\alpha\in\mathbb{R}$) and 
$\widetilde{L}_k = \lambda_k^{-1/2}L^*_k$, we find a GKSL representation of 
$\widetilde{\Ll}$.

Since $[\widetilde{H},\rho]=0$, $\tr(\rho \widetilde{L}_k)=0$ for every $k$ 
and $\{\unit,\widetilde{L}_k\,\mid\,k\geq 1\}$ is clearly a set of linearly
independent elements, we found a special GKSL representation  of $\widetilde{\Ll}$  
by means of the operators $\widetilde{H},\widetilde{L}_k$. Moreover, we have
\[
\rho\widetilde{L}_k=\lambda_k^{-1/2}\rho{L}^*_k
= \lambda_k^{-1/2}\lambda_k^{-1}{L}^*_k \rho= \lambda_k^{-1}\widetilde{L}_k\rho
\]
and, in the same way $\rho\widetilde{L}^*_k\rho^{-1}=\lambda_k^{-1}\widetilde{L}^*_k$. 
Therefore we found a privileged GKSL representation  of $\widetilde{\Ll}$  
by means of the operators $\widetilde{H},\widetilde{L}_k$.
\qed

\section{Quantum detailed balance}\label{sect-Q-DB}

In this section we characterise the generator of a uniformly continuous 
QMS satisfying the quantum detailed balance condition.

\begin{definition}\label{defs-DB} 
A QMS $\mathcal{\T}$ on $\mathcal{B}(\h)$ satisfies the quantum $s$-{\em detailed balance} 
condition ($s$-{\em DB}\,) with respect to a normal faithful invariant state $\rho$, 
if its generator $\Ll$ and the generator $\widetilde{\Ll}$ of the $s$-dual 
semigroup $\widetilde{\T}$ satisfy
\begin{equation}\label{eq-diff-L-Ltilde}
\Ll(a) - \widetilde{\Ll}(a)= 2i\left[K,a\right]
\end{equation}
with a bounded self-adjoint operator $K$ on $\h$ for all $a\in{\mathcal{B}}(\h)$.
\end{definition}

This definition generalises the concept of classical detailed balance discussed 
in Section \ref{sect-class-DB}. Indeed, a classical Markov semigroup $T$ satisfies 
the classical detailed balance condition if and only if $T=\widetilde{T}$, i.e. 
the generators $A$ and $\widetilde{A}$ coincide. 

\begin{lemma}\label{lem}
If $\,\T$ satisfies the quantum $s$-detailed balance condition then $\widetilde{\T}$ 
is a QMS and the self-adjoint operator $K$ in (\ref{eq-diff-L-Ltilde}) commutes with $\rho$.
\end{lemma}

\dimo The identity (\ref{eq-diff-L-Ltilde}) implies that $\widetilde{\Ll}$ 
is conditionally completely positive. Therefore $\widetilde{\T}$ is a QMS.
Moreover, recalling that $\rho$ is an invariant state for both $\T$ and 
$\widetilde{\T}$ by Proposition \ref{propTtilde}, for any 
$a\in{\mathcal{B}}(\h)$, we have then 
\[
0=\tr\left(\rho(\Ll(a)-\widetilde{\Ll}(a))\right)
=2i\tr(\rho[K,a])=2i\tr([\rho,K]a),
\] 
i.e. $[K,\rho\,]=0$. This completes the proof.
\qed\smallskip\\

Notice $[K,\rho\,]=0$ and equation (\ref{eq-diff-L-Ltilde}) imply that the 
linear operator $\Ll^\prime=\Ll - i[K,\cdot]$ is self-adjoint with respect to the 
scalar product $\langle\cdot, \cdot\rangle_0$  on ${\mathcal{B}}(\h)$.

Throughout this section we consider the duality with $s=0$. 

\begin{proposition}\label{prop-L0pmiH}
Given a special GKSL representation of the generator $\Ll$ of a QMS $\T$ 
by means of operators $H, L_k$. Define 
\[
\Ll_0(a)= -\frac{1}{2}\sum_k\left(L_k^*L_k a
-2L_k^* aL_k + a L_k^*L_k \right).
\]
The QMS $\T$ satisfies the quantum $0$-detailed balance condition if and only 
if $\Ll=\Ll_0+i[H,\cdot]$ with $\Ll_0=\widetilde{\Ll}_0$ and $[H,\rho\,]=0$.
\end{proposition}

\dimo Clearly, if $\Ll=\Ll_0+i[H,\cdot]$ with $\Ll_0=\widetilde{\Ll}_0$ 
and $[H,\rho\,]=0$, the QMS $\T$ satisfies the $0$-DB. Indeed, if ${\Ll}_0$ is 
self-adjoint and $H$ commutes with $\rho$, we have $\widetilde{\Ll}=\Ll_0-i[H,\cdot]$. 
Therefore $\Ll(a) - \widetilde{\Ll}(a)=2i[H,a]$.

Conversely, if $\T$ satisfies the $0$-DB condition can find a privileged 
GKSL of $\Ll$ by means of operators $K,M_k$ by Theorem \ref{caratttildeQMS}. 
Note that $K$ commutes with $\rho$ because it is the Hamiltonian in a 
{\em privileged} GKSL representation. On the other hand, the Hamiltonian 
$K$ in a {\em special} GKSL representation is unique up to a scalar 
multiple of the identity by Theorem \ref{th-special-GKSL}, therefore 
we can take $H=K$ and we know that: 1) $H$ commutes with $\rho$, 
2) the operators $L_k$ and $M_k$ define the same map $\Ll_0$.

It follows that $\Ll=\Ll_0+i[H,\cdot]$ and then $\widetilde{\Ll}
=\widetilde{\Ll}_0-i[H,\cdot]$. Moreover, $\T$ satisfies the $0$-DB condition 
so that $\Ll=\widetilde{\Ll}$. It follows that $\Ll_0=\widetilde{\Ll}_0$.
\qed\smallskip\\

We can now characterise generators $\Ll$ of QMS satisfying the $0$-DB condition.

\begin{theorem}\label{th-HL-0-DB} 
A QMS $\T$ satisfies the $0$-detailed balance condition $\Ll-\widetilde{\Ll}=2i[K,\cdot]$ 
if and only if there exists a privileged GKSL representation of $\Ll$, by means of operators 
$H, L_k$, such that:
\begin{enumerate} 
\item \label{it-sdb-1} $H=K+c$ for some $c\in\mathbb{R}$,
\item \label{it-sdb-2} $\lambda^{-1/2}_k {L}_k^* = \sum_j u_{kj}L_j$ for some $\lambda_k>0$ 
and some unitary operator $(u_{kj})_{kj}$ on $\mathsf{k}$.
\end{enumerate} 
In particular both $H$ and $\sum_k L^*_kL_k$ commute with $\rho$.
\end{theorem} 

\dimo If $\T$ satisfies the $0$-DB condition its generator $\Ll$ and 
the generator $\widetilde{\Ll}$ of the dual QMS satisfy $\Ll(a)-i[K,a]= 
\widetilde{\Ll}(a)+i[K,a]$. Let $H, L_k$ be the operators in a privileged 
GKSL representation of $\Ll$. By Theorem \ref{HLtilde}, the operators $\widetilde{H}
=-H-c$ and $\widetilde{L}_k=\lambda_k^{-1/2}L_k^*$ give us a privileged 
GKSL representation of $\widetilde{\Ll}$.

It follows that the operators $H-K, L_k$ and $-H+K-c, \lambda_k^{-1/2}L_k^*$ 
arise in a special GKSL representation of $\Ll(\cdot)-i[H,\cdot]$. Therefore, 
by Theorem \ref{th-special-GKSL}, $H-K=-H+K-c'$ leading us to (\ref{it-sdb-1}) 
and there exists a unitary operator $(u_{kj})_{kj}$ on $\mathsf{k}$ such that 
(\ref{it-sdb-2}) holds. 

Conversely if conditions (\ref{it-sdb-1}) and (\ref{it-sdb-2}) hold, 
writing $\Ll(a)=\Ll_0(a)+i[H,a]$, a straightforward computation shows that 
${\hbox{\rm tr}}(\rho{\widetilde{\Ll}}(a)b)={\hbox{\rm tr}}(\rho a{\Ll}(b))$ 
with $\widetilde{\Ll}(a)=\Ll_0(a)-i[H,a]$. We have then 
${\Ll}(a)-\widetilde{\Ll}(a)=2i[H,a]$ and the $0$-DB condition holds with 
$K=H$.
\qed

\begin{remark}\rm
The proof also shows that we can replace ``there exists a privileged GKSL ..." 
by ``for every privileged GKSL ... "  in Theorem \ref{HLtilde}.
\end{remark}

We conclude this section by showing an example of a QMS $\T$ whose $s$-dual 
semigroup $\widetilde{\T}$ is still a QMS but does not satisfy the 
$s$-detailed balance condition.

\begin{example}\rm\label{tildeQMSnoDB}
We consider $\h=\ell^2(\mathbb{Z}_n,\mathbb{C})$, $n\geq 3$, with the orthonormal 
basis $(e_{{j}})_{j=1,\ldots,n}$, and define 
\[
\Ll(a)=S^*aS-a,
\] 
where $S$ is the unitary shift operator on $\ell^2(\mathbb{Z}_n)$, 
$Se_{{j}}=e_{{j+1}}$ (sum modulo $n$).

The $QMS$ $\T$ generated by $\Ll$ admits $\rho=n^{-1}\unit$ as a 
faithful invariant state because $\Ll_*(\unit)=SS^*-\unit=0$. 
A straightforward computation shows that the dual semigroup $\widetilde{\T}$ 
is the QMS generated by the linear map $\widetilde{\Ll}$ defined by 
$\widetilde{\Ll}(a)=SaS^*-a$.

We now check that $\T$ does not satisfy the $0$-detailed balance condition. 

Letting $H=0$ and $L_1=S$ we find a privileged GKSL representation 
of $\Ll$. Suppose that $\T$ satisfies the $0$-detailed balance condition. 
Then, by Theorem \ref{HLtilde} (\ref{it-sdb-1}), $\Ll=\widetilde{\Ll}$ 
because $K$ is a multiple of the identity operator. This identity, 
however, is not true since
\[
\Ll(|e_2\rangle\langle e_2|)-\widetilde{\Ll}(|e_2\rangle\langle e_2|)
= |e_1\rangle\langle e_1| - |e_3\rangle\langle e_3| \not=0.
\]
Note that the condition $n\ge 3$ is necessary. Indeed, for $n=2$, 
we can easily check that $\Ll=\widetilde{\Ll}$ and the $s$-detailed 
balance condition holds for all $s\in[0,1]$.
\end{example}

\section{Quantum Markov semigroups on $M_2(\mathbb{C})$}\label{sect-QMS-onM2C}

In this section we study in detail the case $\h=\mathbb{C}^2$ and 
$\mathcal{B}(\h)=M_2(\mathbb{C})$. We establish the general form 
of the generator of a QMS $\T$ whose $0$-dual $\widetilde{\T}$ is 
a QMS and show that, in this case, $\T$ satisfies the $0$-detailed 
balance condition.

This can be viewed as a non-commutative counterpart of a well-known 
fact: any $2$-state classical Markov chain satisfies the classical 
detailed balance condition.

We consider, as usual, the basis $\{\sigma_0,\sigma_1,\sigma_2,\sigma_3\}$ 
of $M_2(\mathbb{C})$, where 
\[
\sigma_0=\unit,\ \ \sigma_1=\left(\begin{array}{cc}0&1\\1&0\end{array}\right),\ \ 
\sigma_2=\left(\begin{array}{cc}0&-i\\i&0\end{array}\right),\ \ 
\sigma_3=\left(\begin{array}{cc}1&0\\0&-1\end{array}\right)
\] 
are the Pauli matrices. Any state on $M_2(\mathbb{C})$ has the form 
\[
\frac{1}{2}\left(\sigma_0+u_1\sigma_1+u_2\sigma_2+u_3\sigma_3\right)
\] 
for some vector $(u_1,u_2,u_3)$, in the unit ball of $\mathbb{R}^3$. 
This state is faithful if the vector $(u_1,u_2,u_3)$ belongs to the 
interior of the unit ball, i.e. $u_1^2+u_2^2+u_3^2<1$. After a suitable 
change of coordinates then we can write a faithful state as  
\[
\rho=\left(\begin{array}{cc}
            \nu&0\\0&1-\nu
           \end{array}
     \right)
=\frac{1}{2}\left(\sigma_0+(2\nu-1)\sigma_3\right)
\] 
for some $0<\nu<1$. 

We can now characterise special GKSL representations of the generator 
$\Ll$ of a QMS on $M_2(\mathbb{C})$ in the following way

\begin{lemma}\label{SLR2}
If $L_k=\sum_{j=0}^3z_{kj}\sigma_j$ with $z_{kj}\in\mathbb{C}$, 
$k\in{\mathcal{J}}\subseteq\mathbb{N}$, then 
\[
\Ll(a)=i[H,a]-\frac{1}{2}\sum_{k\in{\mathcal{J}}}
\left(L^*_kL_ka-2L^*_kaL_k+aL^*_kL_k\right)
\] 
is a special GKSL representation of $\Ll$ with respect to $\rho$ if and only if
\begin{enumerate} 
\item $L_k=-(2\nu-1)z_{k3}{\hbox{\em 1\kern-2.8truept l}}
      +\sum_{j=1}^3z_{kj}\sigma_j$ for all $k\in{\mathcal{J}}$, 
\item $\mbox{card}({\mathcal{J}})\leq 3$ and $\{z_k:k\in{\mathcal{J}}\}$ 
(with $z_k=(z_{k1},z_{k2},z_{k3})\,$) is a set of 
linearly independent vectors in $\mathbb{C}^3$.
\end{enumerate} 
\end{lemma}
\dimo 
A simple calculation shows that $\tr(\rho L_k)=2(z_{k0}+(2\nu-1)z_{k3})$
thus, the condition $\tr(\rho L_k)=0$ is equivalent to $z_{k0}=-(2\nu-1)z_{k3}$. 

Finally, $\{\unit,L_k:k\in{\mathcal{J}}\}$ is a set of linearly independent elements 
in $M_2(\mathbb{C})$ if and only if the vectors of coefficients w.r.t. the Pauli matrices 
\[
\{(1,0,0,0),(-(2\nu-1)z_{k3},z_{k1},z_{k2},z_{k3}):k\in{\mathcal{J}}\}
\] 
are linearly independent in $\mathbb{C}^4$; this is clearly equivalent to have 
$\mbox{card}({\mathcal{J}})\leq 3$ and $\{z_k:k\in{\mathcal{J}}\}$ linearly independent 
on $\mathbb{C}^3$, $z_k:=(z_{k1},z_{k2},z_{k3})$.
\qed

\begin{theorem}\label{TtildeQMS2} 
Suppose $\nu\not=1/2$ (i.e. $\rho\not={\hbox{\em 1\kern-2.8truept l}}/2$)
and $\rho$ invariant for $\T$. Then $\tilde{\T}$ is a QMS if and
only if the special Lindblad representation of $\Ll$ has the form
\begin{eqnarray}\label{formaL2x2}
\Ll(a)& = & i[H,a] - \frac{\mid\!\eta\!\mid^2}{2}\left(L^2a-2LaL+aL^2\right) \\
& & -\frac{\mid\!\lambda\!\mid^2}{2}\left(\sigma^-\kern-2truept\sigma^+a
      -2\sigma^-a\sigma^+ +a\sigma^-\kern-2truept\sigma^+\right)
-\frac{\mid\!\mu\!\mid^2}{2}\left(\sigma^+\sigma^-a-2\sigma^+a\sigma^-+a\sigma^+\sigma^-\right),
\end{eqnarray}
where
\begin{eqnarray*}
H&=&v_0\sigma_0+v_3\sigma_3=(v_0+v_3)\sigma^+\sigma^-+(v_0-v_3)\sigma^-\sigma^+,\\
L&=&-(2\nu-1){\hbox{\em 1\kern-2.8truept l}}+\sigma_3=(1-\nu)\sigma^+-\nu\sigma^-,\end{eqnarray*}
$\sigma^+=\sigma_1+i\sigma_2$, $\sigma^-=\sigma_1-i\sigma_2$,
$v_0,v_1\in\mathbb{R}$ and $\lambda,\mu,\eta\in\mathbb{C}$
satisfy
\begin{equation}\label{rhoinv}
{\mid\!\lambda\!\mid^2}/{\mid\!\mu\!\mid^2}={\nu}/{(1-\nu)}.
\end{equation}
\end{theorem}
\dimo
Consider a special GKSL representation 
\[
\Ll(a)=i[H,a]-\frac{1}{2}\sum_{k\in\mathcal{J}}\left(L^*_kL_ka-2L^*_kaL_k+aL^*_kL_k\right)
\]
of $\Ll$ with respect to $\rho$, where $\mathcal{J}\subseteq\{1,2,3\}$, 
$H=\sum_{j=0}^3v_j\sigma_j$ and
\[
L_k=-(2\nu-1)\unit+\sum_{j=1}^3z_{kj}\sigma_j=
\left(\begin{array}{cc}2(1-\nu)z_{k3}& (z_{k1}-iz_{k2})\\
 (z_{k1}+iz_{k2})&-2\nu z_{k3}\end{array}\right),
\]
$\{z_k:k\in{\mathcal{J}}\}$ linearly independent (Lemma \ref{SLR2}). 

We must find $v_j $ and $ z_{kj}  $ such that:
\begin{enumerate}
\item $[H,\rho]=0$;
\item $\rho L_k\rho^{-1}=\lambda_kL_k$ and $\rho L^*_k\rho^{-1}=\lambda_kL^*_k$
for some $\lambda_k>0$;
\item $\rho$ is $\T$-invariant.
\end{enumerate}

$(1)$ Clearly $H$ commutes with $\rho$ if and only if $v_1=v_2=0$, i.e.
\[
H=v_0\unit+v_3\sigma_3=(v_0+v_3)\sigma^+\sigma^-+(v_0-v_3)\sigma^-\sigma^+.
\]

$(2)$ Fix $k\in{\mathcal{J}}$. One can easily check that
\[
\rho L_k\rho^{-1}=
\left(\begin{array}{cc}2(1-\nu)z_{k3}&\frac{\nu}{(1-\nu)}(z_{k1}-iz_{k2})\\
\frac{1-\nu}{\nu}(z_{k1}+iz_{k2})&-2\nu z_{k3}\end{array}\right),
\] 
and, since $\nu\not=1/2$, the identity $\rho L_k\rho^{-1}=\lambda_kL_k$ holds 
if and only if either 
\begin{equation}\label{ce}\left\{\begin{array}{lll}\lambda_k=1\\
z_{k1}-iz_{k2}=0\\
z_{k1}+iz_{k2}=0,\end{array}\right.
\ \ \mbox{or}\ \
\left\{\begin{array}{lll}z_{k3}=0\\
\left(\frac{\nu}{1-\nu}-\lambda_k\right)(z_{k1}-iz_{k2})=0\\
\left(\frac{1-\nu}{\nu}-\lambda_k\right)(z_{k1}+iz_{k2})=0,\end{array}\right.
\end{equation}
In the first case, we get
$L_k=z_{k3}\left(-(2\nu-1)\unit+\sigma_3\right)=
z_{k3}\left((1-\nu)\sigma^+-\nu\sigma^-\right)$; since
$\{L_k:k\in{\mathcal{J}}\}$ is a set of linearly independent elements in
$M_2(\mathbb{C})$, this means that there exists an unique
$k_0\in\mathcal{J}$ such that $\lambda_{k_0}=1$. We can suppose $k_0=3$.\\
Therefore, for $k=1,2$, conditions (\ref{ce}) are equivalent to
\[
\left\{\begin{array}{lll}z_{k3}=0\\
\frac{\nu}{1-\nu}=\lambda_k\\
z_{k1}+iz_{k2}=0\end{array}\right.\
\ \mbox{or}\ \ \
\left\{\begin{array}{lll}z_{k3}=0\\
z_{k1}-iz_{k2}=0\\
\frac{1-\nu}{\nu}=\lambda_k,\end{array}\right.
\]
that is
\[
L_k=\left(\begin{array}{cc}0&-iz_{k2}\\
0&0\end{array}\right)=-iz_{k2}\sigma^+\ \ \mbox{and}\ \
\lambda_k=\frac{\nu}{1-\nu},
\]
or
\[
L_k=\left(\begin{array}{cc}0&0\\
iz_{k2}&0\end{array}\right)=iz_{k2}\sigma^-\ \ \mbox{and}\ \
\lambda_k=\frac{1-\nu}{\nu},
\] 
so that we have $L_1=-iz_{12}\sigma^+=\lambda\sigma^+$ and
$L_2=iz_{22}\sigma^-=\mu\sigma^-$, with $\lambda_1=\nu/(1-\nu)$
and $\lambda_2=\lambda_1^{-1}$.

Moreover, with this choice of $L_1, L_2$ and $L_3$, the equalities
$\rho L^*_k\rho^{-1}=\lambda_k^{-1}L^*_k$ are automatically
satisfied.
 
$(3)$ Since $H,L_3$ and $\rho$ commute, $\rho$ is $\T$-invariant if
and only if
\begin{eqnarray*}
0&=&\frac{1}{2}\sum_{k=1}^2\left(L^*_kL_k\rho-2L_k\rho L^*_k+\rho
L^*_kL_k\right)\\
&=&\frac{1}{2}\sum_{k=1}^2\left(L^*_kL_k\rho-2L_k(\rho
L^*_k\rho^{-1})\rho+(\rho L^*_k\rho^{-1})(\rho
L_k\rho^{-1})\rho\right)\\
&=&\sum_{k=1}^2\left(L^*_kL_k\rho-\lambda^{-1}_kL_kL^*_k\right),
\end{eqnarray*}
that is
$$\frac{\mid\!\lambda\!\mid^2}{\mid\!\mu\!\mid^2}=
\frac{\mid\!z_{12}\!\mid^2}{\mid\!z_{22}\!\mid^2}=\frac{\nu}{1-\nu}.$$
This concludes the proof.

\qed

\begin{theorem} Suppose $\nu\not=1/2$.\\ If $\widetilde{\T}$ is a QMS, then $\T$ satisfies detailed
balance.
\end{theorem}
\dimo
By Theorem \ref{TtildeQMS2} there exists a privileged GKSL representation of $\mathcal{L}$ with
\begin{equation} \label{rellamdamunu}
\left\{\begin{array}{ccc}L_1=\eta L\\ L_2=\lambda\sigma^+\\ L_3=\mu\sigma^-\end{array}\right.\quad\quad
\left\{\begin{array}{ccc}\lambda_1=1\\ \lambda_2=\frac{\nu}{1-\nu}\\ \lambda_3=\lambda_2^{-1}\end{array}\right.\quad\quad\mbox{and}\quad\quad
\frac{\mid\!\lambda\!\mid^2}{\mid\!\mu\!\mid^2}=\frac{\nu}{1-\nu}.
\end{equation}
Therefore,
\begin{equation}\label{matrice}
\left(\begin{array}{ccc}\sqrt{\lambda_1^{-1}}L^*_1\\ \sqrt{\lambda_2^{-1}}L^*_2\\
\sqrt{\lambda_3^{-1}}L^*_3\end{array}\right)=\left(\begin{array}{ccc}\overline{\eta} L\\ \sqrt{\frac{1-\nu}{\nu}}\overline{\lambda}\sigma^-\\
\sqrt{\frac{\nu}{1-\nu}}\overline{\mu}\sigma^+\end{array}\right)=
\left(\begin{array}{ccc}\eta/\overline{\eta}&0&0\\ 0& 0&\sqrt{\frac{1-\nu}{\nu}}\frac{\overline{\lambda}}{\mu}\\
0& \sqrt{\frac{1-\nu}{\nu}}\frac{\overline{\mu}}{\lambda} \end{array}\right)
\left(\begin{array}{ccc}L_1\\ L_2\\ L_3\end{array}\right)
\end{equation}
and $$\left(\begin{array}{ccc}\eta/\overline{\eta}&0&0\\ 0& 0&\sqrt{\frac{1-\nu}{\nu}}\frac{\overline{\lambda}}{\mu}\\
0& \sqrt{\frac{1-\nu}{\nu}}\frac{\overline{\mu}}{\lambda} \end{array}\right)$$ is unitary thanks to (\ref{rellamdamunu}).
\qed

\section{The symmetric dual semigroup and detailed balance condition}\label{sect-symmBD}

We now study the $s$-dual semigroup and the quantum $s$-detailed balance 
condition for $s=1/2$. In this case we call ${\T}^\prime$ the {\em symmetric 
dual semigroup} of $\T$ and call {\em symmetric detailed balance} condition 
the $1/2$-detailed balance condition. 

By Proposition \ref{propTtilde}, the symmetric dual semigroup of $\T$ is 
defined by
\begin{equation} \label{defsimm}
\rho^{1/2}{\mathcal{\T}}_t^\prime(a)\rho^{1/2}=
\mathcal{T}_{*t}(\rho^{1/2}a\rho^{1/2}),
\end{equation} 
so that 
$$
\mathcal{\T}_t^\prime(a)\supseteq\rho^{-1/2} \mathcal{T}_{*t}(\rho^{1/2}a\rho^{1/2})\rho^{-1/2}
$$
for all $a\in{\mathcal{B}}(\h)$.
The name symmetric is then justified by 
the left-right symmetry of multiplication by $\rho^{1/2}$ and $\rho^{-1/2}$. 
Equation (\ref{defsimm}) ensures that any map ${\T}_t^\prime$ is completely 
positive, contrary to the case $s=0$ (Example \ref{tildenoQMS}). 
Therefore the symmetric dual semigroup ${\T}^\prime$ is always a QMS with 
generator given by (Proposition \ref{prop-s-db-generators})
\[
\rho^{1/2}{\Ll}^\prime(a)\rho^{1/2}=\Ll_*(\rho^{1/2}a\rho^{1/2}).
\]

The relationship between dual semigroups $\widetilde{\T}$ and $\T^\prime$ 
is described by the following

\begin{theorem}\label{0duale=1/2} 
The $0$-dual $\widetilde{\T}$ and the symmetric dual $\T^\prime$ of a QMS 
$\T$ coincide if and only if each map $\T_t$ commutes with $\sigma_{-i}$.
\end{theorem} 

\dimo
If $\widetilde{\T}={\T}^\prime$, then $\widetilde{\T}$ is a QMS; 
hence, by Theorem \ref{th-dual-QMS-impl-commautmod}, 
$\T$ commutes with the modular automorphism $\sigma_ {-i}$. 

On the other hand, we showed in the proof of Theorem \ref{th-dual-QMS-impl-commautmod} 
that the commutation between $\T$ and $\sigma_{-i}$ implies 
$\widetilde{\mathcal{T}}_t(a)
=\rho^{-1/2}\mathcal{T}_{*t}(\rho^{1/2}a\rho^{1/2})\rho^{-1/2}$ for all
$a\in{\mathcal{B}}(\h),\ t\geq 0$, and then $\widetilde{\mathcal{T}}
={\mathcal{T}}^\prime$.
\qed\smallskip\\

We now establish the relationship between the generator $\Ll$ of a QMS and the 
generator ${\Ll}^\prime$ of the symmetric dual semigroup.

\begin{theorem}\label{dualesimm} 
For all special GKSL representation $\Ll(a)=G^*a+\sum_kL^*_kaL_k+aG$ of $\Ll$ 
there exists a special GKSL representation of $\Ll^\prime$ by means of operators $G^\prime,L^\prime_k$ 
such that:
\begin{enumerate} 
\item\label{cnd-symm-DB-1inf} 
     $G^\prime\rho^{1/2}=\rho^{1/2}G^*+ic\rho^{1/2}$ for some $c\in\mathbb{R}$,
\item\label{cnd-symm-DB-2inf} 
     $L^\prime_k\rho^{1/2}=\rho^{1/2}L^*_k$
\end{enumerate}
\end{theorem}

\dimo Since $\mathcal{T}^\prime$ is a uniformly continuous QMS, its
generator $\Ll^\prime$ admits a special GKSL representation,
$\mathcal{L}^\prime(a)=G^{\prime *}a+\sum_kL^{\prime *}_kaL^\prime_k+aG^\prime$.
Moreover, by Proposition \ref{Gcomm} we have
$G^\prime\rho^{1/2}=\rho^{1/2}G^*+ic$, $c\in\mathbb{R}$, and so the relation
$\rho^{1/2}\Ll^\prime(a)\rho^{1/2}=\Ll_*(\rho^{1/2}a\rho^{1/2})$
implies
\begin{equation}\label{partecp}
\sum_k\rho^{1/2}L^{^\prime *}_kaL^\prime_k\rho^{1/2}=
\sum_kL_k\rho^{1/2}a\rho^{1/2}L^*_k.
\end{equation}
Define
$$X(x\otimes\unit_{\mathsf{k}^\prime})L^\prime\rho^{1/2}u
=(x\otimes\unit_{\mathsf{k}})(\rho^{1/2}\otimes\unit_\mathsf{k})L^*u$$ 
for all $x\in\B$ and $u\in\h$, where $L:\h\rightarrow\h\otimes \mathsf{k}$, 
$Lu=\sum_kL_ku\otimes f_k$, $L^\prime:\h\rightarrow\h\otimes \mathsf{k}^\prime$, 
$L^\prime u=\sum_kL^\prime_ku\otimes f^\prime_k$, $(f_k)_k$ and $(f^\prime_k)_k$ 
orthonormal basis of $\mathsf{k}$ and $\mathsf{k}^\prime$ respectively.
Thus, by (\ref{partecp}),
$$
\langle X(x\otimes\unit_{\mathsf{k}^\prime})L^\prime\rho^{1/2} u,X(y\otimes\unit_{\mathsf{k}^\prime})L^\prime \rho^{1/2}v\rangle
=\sum_k\langle u,\rho^{1/2}L^{\prime *}_k x^*yL^{\prime}_k \rho^{1/2}v\rangle
$$
$$
=\langle(x\otimes\unit_{\mathsf{k}^\prime})
L^\prime \rho^{1/2}u,(y\otimes\unit_{\mathsf{k}^\prime})L^\prime \rho^{1/2}v\rangle
$$
for all $x,y\in\B$ and $u,v\in\h$, i.e. $X$ preserves the scalar product.
Therefore, since the set $\{(x\otimes\unit_{\mathsf{k}^\prime})L^\prime\rho^{1/2} u\,\mid\,x\in\B,\ u\in\h\}$ is total in $\h\otimes\mathsf{k}^\prime$ (for $\rho^{1/2}(\h)$ is dense in $\h$ and Theorem \ref{DLMconL} holds), 
we can extend $X$ to an
unitary operator from $\h\otimes \mathsf{k}^\prime$ to $\h\otimes
\mathsf{k}$.
As a consequence we have $X^*X=\unit_{\h\otimes \mathsf{k}^\prime}$.

Moreover, since
$X(y\otimes\unit_{\mathsf{k}^\prime})=(y\otimes\unit_{\mathsf{k}^\prime})X$
for all $y\in\B$, we can conclude that $X=\unit_\h\otimes Y$ for
some unitary map
$Y:\mathsf{k}^\prime\rightarrow\mathsf{k}^\prime$. 

The definition of $X$ implies then 
$$ (\rho^{1/2}\otimes\unit_{\mathsf{k}})L^*=XL^\prime\rho^{1/2}=(\unit_\h\otimes Y)L^\prime\rho^{1/2}.$$

This means that, by substituting $L^\prime$ by $(\unit_\h\otimes Y)L^\prime$, or more precisely $L^\prime_k$ by $\sum_lu_{kl}L^\prime_l$ for all $k$, we have 
$$
\rho^{1/2}L^*_k=L^\prime_k\rho^{1/2}.
$$

Since $\tr(\rho {L^\prime}_k)= \tr(\rho L_k^*)=0$ and, from 
${\Ll}^\prime(\unit)=0$ (Proposition \ref{propTtilde}), ${G^\prime}^*+ 
{G^\prime}=\sum_k {L^\prime}^*_k {L^\prime}_k$,  
the properties of a special GKSL representation follow. 
\qed\smallskip\\

Contrary to what happens in the case $s=0$, the operators $G, {G}^\prime$ 
may not commute with $\rho$, as the following example shows. 

\begin{example}\rm\label{example-H-no-comm-rho-dual}
Fix a faithful state $\rho=(\unit+(2\nu-1)\sigma_3)/2$ on $M_2(\mathbb{C})$ 
with $\nu\in]0,1[$, $\nu\not=1/2$ and consider the semigroup on $M_2(\mathbb{C})$ 
generated by 
\[
\Ll(a)=i[H,a]-\frac{1}{2}\left(L^*La-2L^*aL+aL^*L\right)
\]
with $H=\Omega\sigma_1, \, L=(1-2\nu)\unit+ir\sigma_1+s\sigma_2+\sigma_3$ and 
$\Omega, r,s\in\mathbb{R}$, $\Omega\not=0$. Clearly $\Ll$ is represented 
in a special GKSL form with respect to the faithful state $\rho$ and 
$H$ does not commute with $\rho$.

We now show that $\rho$ is an invariant state for the QMS generated by $\Ll$ 
for a special choice of the constants $\Omega, r,s$ and so we find the desired 
example. 

A long but straightforward computation shows that, 
if we choose $r,s$ satisfying
\begin{equation}\label{eq-ri}
2\nu=(r-s)^2/(r^2+s^2),
\end{equation}
then we find
$\Ll_*(\rho) = \left((-4\nu^2+4\nu+1)r+(2\nu-1)(s-\Omega)\right)\sigma_2$.
%Therefore, if we take $\Omega,r,s$ satisfying (\ref{eq-ri}) and 
%\begin{equation}\label{eq-si}
%s=\Omega+r\left(4\nu^2-4\nu-1\right)/(2\nu-1).
%\end{equation}
It is now a simple exercise to show that for all fixed $\nu$ and $\Omega\not=0$ 
there exist $r,s$ satisfying (\ref{eq-ri}) and 
\begin{equation}\label{eq-si}
s=\Omega+r\left(4\nu^2-4\nu-1\right)/(2\nu-1).
\end{equation} A little computation yields
\begin{equation}\label{eq-erreesse}
s=\frac{\pm \Omega}{2\sqrt{\nu(1-\nu)}}, \qquad 
r= \frac{\pm \Omega (1-2\nu)}{2 \sqrt{\nu(1-\nu)}
   \left(1\pm 2\sqrt{\nu(1-\nu)}\right)}
\end{equation}
($\pm$ are all $+$ or all $-$).
With this choice of $r$ and $s$ the state $\rho$ is invariant. 
\end{example}

The $0$-detailed balance is stronger than the symmetric detailed 
balance (see also \cite{Cip} Th. 6.6 p.296).

\begin{proposition}\label{1/2impl0} 
If $\,\T$ satisfies the $0$-detailed balance, then it also fulfills the 
symmetric detailed balance. Moreover, these conditions are equivalent 
if and only if the $0$-dual $\widetilde{\T}$ is a QMS.
\end{proposition} 

\dimo Suppose that $\widetilde{\T}$ is a QMS. As we showed in 
the proof of Theorem \ref{th-dual-QMS-impl-commautmod}, $\widetilde{{\T}}_t(a)
=\rho^{-1/2}{\T}_{*t}(\rho^{1/2}a\rho^{1/2})\rho^{-1/2}$. 
Then $\widetilde{{\T}}={{\T}}^\prime$ by (\ref{defsimm}), as a 
consequence $\widetilde{\Ll}=\Ll^\prime$, i.e. $\Ll-\widetilde{\Ll}=\Ll-\Ll^\prime$,  
and both detailed balance conditions are equivalent. 

On the other hand, if $\T$ satisfies the $0$-detailed balance, then 
$\widetilde{\T}$ is a QMS. Therefore $\widetilde{\T}=\T^\prime$ 
and $\T$ also fulfills the symmetric detailed balance condition. 
\qed\smallskip\\

We end this section by finding the relationships between the operators 
$H,L_k$ in a special GKSL representation of the generator of a QMS 
satisfying symmetric detailed balance.

\begin{theorem}\label{th-gen-symm-DB} 
A QMS $\T$ satisfies the symmetric detailed balance condition 
$\Ll-\Ll'=2i[K,\cdot]$ if and only if there exists a special GKSL 
representation of the generator $\Ll$ by means of operators $H,L_k$ 
such that, letting $2G=-\sum_kL^*_kL_k - 2iH$, we have:
\begin{enumerate} 
\item\label{cnd-symm-DB-1} 
     $G\rho^{1/2}=\rho^{1/2}G^* -\left(2i K +ic\right)\rho^{1/2}$ for some $c\in\mathbb{R}$,
\item\label{cnd-symm-DB-2} 
     $\rho^{1/2}L^*_k=\sum_\ell u_{k\ell}L_\ell\rho^{1/2}$, 
for all $k$, for some unitary matrix $(u_{k\ell})_{k\ell}$.
\end{enumerate}
\end{theorem} 

\dimo Choose a special GKSL representation of $\Ll$ by means 
of operators $H$, $L_k$. 
Theorem \ref{dualesimm} allows us to write the dual $\Ll^\prime$ in a
special GKSL representation by means of operators $H^\prime$, $L^\prime_k$
with $H^\prime=(G^{\prime *}-G^{\prime})/(2i)$, 
\begin{equation}\label{eq-Gprime-Lprime}
G^\prime\rho^{1/2}=\rho^{1/2}G^*,\quad\quad L^\prime_k\rho^{1/2}=\rho^{1/2}L^*_k.
\end{equation}

Suppose first that $\T$ satisfies the symmetric detailed balance condition. 
Then $\Ll-i[K,\cdot] = \Ll^\prime + i[K, \cdot]$ and $K$ commutes with $\rho$ 
by Lemma \ref{lem}. Comparing the special GKSL representations of $\Ll-i[K,\cdot]$ 
and $\Ll^\prime + i[K, \cdot]$, by Theorem \ref{th-special-GKSL} and 
Remark \ref{rem-G-unique} we find
\[
G + i K= G^\prime - i K +ic, \quad  
L^\prime_k = \sum_j u_{kj}L_j,
\]
for some unitary matrix $(u_{kj})_{kj}$ and some $c\in\mathbb{R}$. This, together 
with (\ref{eq-Gprime-Lprime}) implies that conditions (\ref{cnd-symm-DB-1}) and 
(\ref{cnd-symm-DB-2}) hold.

Conversely, notice that the dual $\Ll^\prime$ 
admits the special GKSL representation
\[
\Ll^\prime(a) = G^{\prime*}a + \sum_k L^{\prime*}_kaL^{\prime}_k + a G^\prime.
\] 
Therefore, if conditions (\ref{cnd-symm-DB-1}) and (\ref{cnd-symm-DB-2}) are 
satisfied, by (\ref{eq-Gprime-Lprime}), we have
$$
G^{\prime}\rho^{1/2}=\rho^{1/2}G^*=(G+2iK)\rho^{1/2},
$$
so that $G^{\prime}=G+2iK$ and then
\begin{eqnarray*}
G^{\prime *}a + aG^\prime
& = & (G^*-2iK)a + a(G+2iK) = G^*a + aG - 2i [K,a] \\
\sum_k L^{\prime *}_k a L^{\prime}_k 
& = & \sum_{k,j,m}\bar{u}_{kj}L^*_j a u_{km}L_m 
=\sum_{j,m}\left(\sum_k\bar{u}_{kj} u_{km}\right) L^*_j a L_m 
=\sum_k L^{*}_k a L_k. 
\end{eqnarray*}
It follows that $\Ll^\prime(a)=\Ll(a) - 2i[K,a]$ and the symmetric detailed 
balance condition holds. 
\qed\smallskip\\

The Hamiltonian $H$ in a special GKSL representation of the generator of a 
QMS satisfying the {\em symmetric} detailed balance condition does not need 
to commute with the invariant state $\rho$ (as in the case of $0$-detailed 
balance) as shows the following 

\begin{example}\rm\label{example-H-no-comm-rho-symm-DB}
Let $\Ll$ be the generator described in Example \ref{example-H-no-comm-rho-dual} 
and let $\Ll^\prime$ be its symmetric dual. The linear map  
$\mathcal{K}=(\Ll+\Ll^\prime)/2$ is clearly the generator of a QMS.
Moreover, $\rho$ is an invariant state for $\mathcal{K}$ because it is an 
invariant state for $\Ll$ and $\Ll^\prime$ by Proposition \ref{propTtilde}. 
$\mathcal{K}$ satisfies the symmetric detailed balance condition 
by its definition.

The special GKSL representation of $\Ll$ by means of operators $H, L$ as in 
Example \ref{example-H-no-comm-rho-dual} yields a special GSKL representation 
of $\Ll^\prime$ choosing $L^\prime=\rho^{1/2}L^*\rho^{-1/2}$ and 
$H^\prime=(G^{\prime *} - G^\prime)/(2i)$ with $G^\prime=\rho^{1/2}G^*\rho^{-1/2}$ 
and $2G=-L^*L-iH$. Putting 
%$M_1=L/\sqrt{2}$, $M_2=\rho^{1/2}L^*\rho^{-1/2}/\sqrt{2}$
\begin{eqnarray*}
M_1 = L/{\sqrt{2}}, & & \qquad\qquad M_2= \rho^{1/2}L^*\rho^{-1/2}/{\sqrt{2}}, \\
F=(G+G^\prime)/2, & & \quad  F_0 = (F+F^*)/2, \quad
K = (F^*-F)/(2i)
\end{eqnarray*}
we have $2F_0=M^*_1M_1+M^*_2M_2$ and a special GKSL representation of 
$\mathcal{K}$ by means of operators $K,M_j$. 

We now check that $K$ does not commute with $\rho$. To this end it suffices 
that $K$ is not linearly independent of $\sigma_j$ for $j=1$ or $j=2$, namely 
$\tr(\sigma_j K)\not=0$. But
\[
2\tr(\sigma_j K) = 2\Im\tr(\sigma_j F) = \Im\tr(\sigma_j (G+G^\prime))
= \Im\tr(\sigma_j G)+\Im\tr(\rho^{-1/2}\sigma_j\rho^{1/2}G^*),
\]
where, defining $r$ and $s$ as in (\ref{eq-erreesse}) with 
$-$ signs and computing $(2\nu-1)s+r=(2\nu-1)\Omega/(1-2\sqrt{\nu(1-\nu)})$ we have
\[
G=-\frac{(1-2\nu)^2 + 1+s^2+r^2}{2}\unit-i\Omega\sigma_1
+\frac{(2\nu-1)\Omega}{1-2\sqrt{\nu(1-\nu)}}\sigma_2+((2\nu-1)-rs)\sigma_3.
\]
Another straightforward computation yields 
\[
\rho^{1/2}=\frac{\kappa }{2}\left(\unit + \frac{2\nu-1}{\kappa^2}\sigma_3\right), \quad 
\rho^{-1/2}=\frac{2}{\kappa}\,\frac{1}{1-\kappa^{-4}(2\nu-1)^2}\left(\unit - \frac{2\nu-1}{\kappa^2}\sigma_3\right) 
\]
where $\kappa:=\sqrt{1+2\sqrt{\nu(1-\nu)}}=\sqrt{\nu}+\sqrt{1-\nu}$  and 
\[
\rho^{-1/2}\sigma_1\rho^{1/2} 
= \frac{1}{2\sqrt{\nu(1-\nu)}}\left(\sigma_1-i(2\nu-1)\sigma_2\right).
\]
It follows that $2\tr(\sigma_1 K) =-2\Omega$.
Therefore we find $\tr(\sigma_1 K)\not=0$ for all $\nu\in]0,1[$ with $\nu\not=1/2$ 
and $\Omega\not=0$.
\end{example}  

\section{Case $s\in(0,1/2)\cup(1/2,1)$}
We conclude the discussion on the $s$-dual semigroup by considering $s\in(0,1/2)\cup(1/2,1)$. In this framework, we show that
$\widetilde{\mathcal{T}}^{(s)}$ is a QMS if and only if the
$0$-dual semigroup is a QMS and, in this case, they coincide.
Therefore, it is enough to study the case $s=0$. \\

\begin{proposition} The following facts are equivalent:
\begin{enumerate}\item $\widetilde{\mathcal{T}}^{(s)}$ is a QMS;
\item $\widetilde{\mathcal{T}}^{(0)}$ is a QMS.
\end{enumerate}
Moreover, if the above conditions hold, then
$\widetilde{\mathcal{T}}^{(s)}=\widetilde{\mathcal{T}}^{(0)}$.
\end{proposition}

\dimo
$1\Rightarrow 2$. Since $\widetilde{\mathcal{T}}^{(s)_t}$ and $\mathcal{T}_{*t}$ are $*$-maps, by the second formula (\ref{eq-L-Ltilde}) and the same formula 
taking the adjoint we have
$$
\rho^s\mathcal{T}_t(a)\rho^{1-s}=\widetilde{\mathcal{T}}^{(s)}_t(\rho^sa\rho^{1-s})\ \ \mbox{and}\ \ \rho^{1-s}\mathcal{T}_t(a)\rho^{s}=\widetilde{\mathcal{T}}^{(s)}_t(\rho^{1-s}a\rho^{s}).
$$
Therefore, given $a\in\anal$, we get
$$
\rho^s\mathcal{T}_t(a)\rho^{1-s}=\widetilde{\mathcal{T}}^{(s)}_t(\rho^{1-s}(\rho^{2s-1}a\rho^{1-2s})\rho^{s})=\rho^{1-s}\mathcal{T}_t(\rho^{2s-1}a\rho^{1-2s})\rho^{s}
$$
and then
$$
\mathcal{T}_t(a)=\rho^{1-2s}\mathcal{T}_{t}(\rho^{2s-1}a\rho^{1-2s})\rho^{2s-1}
$$
i.e. any $\mathcal{T}_t$ commutes with $\sigma_{-i(2s-1)}$.\\ This
means that the contraction semigroup $(\hat{\mathcal{T}}_t)$
defined on $L^2(\h)$ by
$$\hat{\mathcal{T}}_t(a\rho^{1/2})=\mathcal{T}_t(a)\rho^{1/2}$$
commutes with $\Delta^{1-2s}$ and then, by spectral calculus, it
also commutes with $\rho$; it follows that $\mathcal{T}_t$
commutes with the modular automorphism $\sigma_{-i}$ and so $\wT$
is a QMS by Theorem \ref{th-dual-QMS-impl-commautmod}.
\smallskip\\
$2\Rightarrow 1$. If $\wT$ is a QMS, by Theorem \ref{caratttildeQMS} there exists a privileged GKSL representation of $\mathcal{L}$ by means of operators $H$ and $L_k$ such that 
$$
\Delta (L_k\rho^{1/2})=\rho L_k\rho^{-1/2}=\lambda_kL_k\rho^{1/2}\ \ \mbox{and}\ \ \Delta (L^*_k\rho^{1/2})=\rho L^*_k\rho^{-1/2}=\lambda^{-1}_kL^*_k\rho^{1/2}.
$$
It follows by spectral calculus that 
\begin{eqnarray*}
\rho^\alpha L_k\rho^{-\alpha}\rho^{1/2}&=&\Delta^\alpha (L_k\rho^{1/2})=\lambda^\alpha_kL_k\rho^{1/2}\\
\rho^\alpha L_k^*\rho^{-\alpha}\rho^{1/2}&=&\Delta^\alpha (L_k^*\rho^{1/2})=\lambda^{-\alpha}_kL_k^*\rho^{1/2}
\end{eqnarray*}
for all $\alpha\not=0$.\\
Therefore, since $H$ and $\rho$ commute, we have
\begin{eqnarray*}
\wLs(a)&=&\rho^{s-1}\mathcal{L}_{*}(\rho^{1-s}a\rho^s)\rho^{-s}=-i\rho^{s-1}[H, \rho^{1-s}a\rho^s]\rho^{-s}\\
&-&\frac{1}{2}\sum_k\left(\rho^{s-1}L^*_kL_k\rho^{1-s}a-2\rho^{s-1}L_k\rho^{1-s}a\rho^{s}L^*_k\rho^{-s}
+a\rho^{s}L^*_kL_k\rho^{-s}\right)\\ &=&-i[H,a]-\frac{1}{2}\sum_k\left(L^*_kL_ka-2\lambda^{-1}_kL_kaL^*_k+aL^*_kL_k\right)
\end{eqnarray*} 
for all $a\in\anal$; since $-i[H,a]-\frac{1}{2}\sum_k\left(L^*_kL_ka-2\lambda^{-1}_kL_kaL^*_k+aL^*_kL_k\right)=\wL(a)$ by Theorem \ref{HLtilde} and $\anal$ is $\sigma$-weakly dense in $\B$, the above equality means $\wLs=\wL$, so that $\wTs$ is a QMS and it coincides with $\wT$.
\qed

\section*{Acknowledgment}
The financial support from the MIUR PRIN 2006-2007 project ``Quantum Markov 
Semigroups and Quantum Stochastic Differential Equations'' is gratefully 
acknowledged.


\vspace*{8pt}

\begin{thebibliography}{99}

\bibitem{AcMo} 
{\textsc L.\ Accardi, A.\ Mohari}, Time Reflected Markov Processes. {\it Infinite Dim. 
Anal. Quantum Probab. Related Topics}, {\bf 2}, 397-426 (1999).

\bibitem{alicki}  {\textsc R.\ Alicki}, On the detailed balance condition for 
non-Hamiltonian systems.  {\it Rep. Math. Phys.}, {\bf 10}, 249-258, (1976).

\bibitem{AlickiLendi}
R.~Alicki, K.~Lendi, {\it Quantum Dynamical Semigroups and 
Applications}, Lect. Notes Phys. 286, Springer-Verlag, 1987.

\bibitem{benfatto}    {\textsc G.\ Benfatto, C.\ D.\ D'Antoni, F.\ Nicol\`{o}, G.\ C.\ Rossi}, 
{\it Introduzione alla teoria delle algebre di von Neumann e teorema di Tomita-Takesaki},  
Quaderni del consiglio nazionale delle ricerche, 1975.

\bibitem{Cip}
F.~Cipriani, ``Dirichlet Forms and Markovian Semigroups on Standard Forms of 
von Neumann Algebras,'' {\it J. Funct. Anal.}, {\bf 147}, 259-300 (1997). 

\bibitem{olabra}    {\textsc O.\ Bratteli and D.\ W.\ Robinson}, 
{\it Operator algebras an quantum statistical mechanics I},  
Springer-Verlag, Berlin-Heidelberg-New York 1979.

\bibitem{FagQue} 
{\textsc F.\ Fagnola, R.\ Quezada}, Two-Photon Absorption and Emission Process. 
{\it Infinite Dim. Anal. Quantum Probab. Related Topics}, {\bf 8}, 573-592, (2005).

\bibitem{FrGo}    
{\textsc A. Frigerio, V. Gorini}, Markov dilations and quantum detailed balance. 
{\it Comm. Math. Phys.} {\bf 93} (1984), no. 4, 517--532.

\bibitem{GoLi} 
{\textsc S.\ Goldstein, J.M.\ Lindsay}, 
KMS symmetric semigroups,  {\it Math. Z.}, 
{\bf 219}, 591--608, (1995).

\bibitem{kossa}  
{\textsc A.\ Kossakowski, A.\ Frigerio, V.\ Gorini and M.\ Verri}, 
Quantum detailed balance and KMS condition,  {\it Comm. Math. Phys.}, 
{\bf 57}, 97-110, (1977).

\bibitem{maje}    {\textsc W.\ A.\ Majewski and R.\ F.\ Streater}, Detailed balance 
and quantum dynamical maps.  {\it J. Phys. A: Math. Gen.}, {\bf 31}, 7981-7995, (1998).

\bibitem{part}    {\textsc K.\ R.\ Parthasarathy}, {\it An introduction to 
quantum stochastic calculus}, Volume 85 of {\it Monographs in Mathematics},  
Birkh\"auser-Verlag, Basel-Boston-Berlin 1992. 


\end{thebibliography}
\end{document}